\documentclass[12pt]{article}
\pdfoutput=1 % USING PDFLaTeX

\usepackage[all,knot]{xy}
\usepackage{graphics,epsfig}
\usepackage{hyperref}   
\usepackage{rotate}
\usepackage{amsfonts,amssymb,amsmath}

\title{\MM\ Gravity and Cartan Geometry}
\author{Derek K.\ Wise
\\ Department of Mathematics
\\ University of California
\\ Davis, CA 95616, USA
\\
\\ email: derek@math.ucdavis.edu
}
\date{}

\newcommand{\GR}{general relativity}
\newcommand{\MM}{MacDowell--Mansouri}

\newcommand{\bfm}{\boldmath \bf}

\newcommand{\half}{\frac{1}{2}}

\newcommand{\xa}{\alpha}
\newcommand{\xb}{\beta}
\newcommand{\xg}{\gamma}
\newcommand{\xd}{\delta}
\newcommand{\xe}{\epsilon}
\newcommand{\om}{\omega}

\newcommand{\R}{{\mathbb R}}

\def\tr {{\rm tr\,}}
\def\del {\nabla}
\newcommand{\maps}{\colon}
\def\ker {{\rm ker}}

\def\stackto #1 { \, {\stackrel{#1}{\longrightarrow}}\, }
\def\stackTo #1 { {\stackrel{#1}{\Longrightarrow}} }

\def\iso{\cong}

\newcommand{\Vect}{{\rm Vect}}

\newcommand{\frakg}{\mathfrak{g}} %
\newcommand{\frakh}{\mathfrak{h}} %

\newcommand{\Ad}{{\rm Ad}}
\newcommand{\SO}{{\rm SO}}
\newcommand{\so}{\mathfrak{so}}

\newcommand{\ISO}{{\rm ISO}}
\newcommand{\Iso}{\mathfrak{iso}}

\newcommand{\g}{\mathfrak{g}}
\newcommand{\h}{\mathfrak{h}}
\newcommand{\p}{\mathfrak{p}}  % transvections

\def\subgp {\subset}

\newcommand{\Hyp}{\mathrm{H}} % Hyperbolic 

\newcommand{\MdS}{M_{\text{\rm dS}}} % de Sitter spacetime

\newcommand{\fake}{{\cal T}}  % The fake tangent bundle

\newcounter{letter} \newcounter{numeral} \newcounter{Numeral}

\newtheorem{defn}{Definition}

\newcommand{\bt}{b} % transvection part of B field 
\newcommand{\bs}{\beta} % stabilizer, or "rotational", part of B field

\newcommand{\arxiv}[1]{\href{http://arxiv.org/abs/#1}{arXiv:\nolinkurl{#1}}}
\newcommand{\narxiv}[2]{\href{http://arxiv.org/abs/#1}{arXiv:\nolinkurl{#1 [#2]}}}

\newcommand{\beq}{\begin{equation}}
\newcommand{\eeq}{\end{equation}}

\begin{document}

\maketitle
\thispagestyle{empty}
\abstract{
The geometric content of the \MM\ formulation of general relativity is best understood in terms of Cartan geometry.  In particular, Cartan geometry gives clear geometric meaning to the \MM\ trick of combining the Levi-Civita connection and coframe field, or soldering form, into a single physical field.  The Cartan perspective allows us to view physical spacetime as tangentially approximated by an arbitrary homogeneous `model spacetime', including not only the flat Minkowski model, as is implicitly used in standard general relativity, but also de Sitter, anti de Sitter, or other models.   A `Cartan connection' gives a prescription for parallel transport from one `tangent model spacetime' to another, along any path, giving a natural interpretation of the \MM\ connection as `rolling' the model spacetime along physical spacetime.  I explain Cartan geometry, and `Cartan gauge theory', in which the gauge field is replaced by a Cartan connection.  In particular, I discuss \MM\ gravity, as well as its more recent reformulation in terms of $BF$ theory, in the context of Cartan connections.  
}

\newpage

\section{Introduction}
\label{sec:intro}

The geometric content of standard general relativity, as described by Riemannian geometry, has long been well understood.  In the late 1970s, MacDowell and Mansouri introduced a new approach, based on broken symmetry in a type of gauge theory \cite{MM}, which has since been influential a wide range of gravitational theory. 
However, despite their title ``Unified geometric theory of gravity and supergravity", the geometric meaning of the \MM\ approach is relatively obscure.  In the original paper, and in much of the work based on it, the technique seems like an unmotivated `trick' that just happens to reproduce the equations of general relativity. 

In fact, the secret to a geometric interpretation of their work had been around in some form for over 50 years by the time MacDowell and Mansouri introduced their theory.  The geometric foundations had been laid in the 1920s by \'Elie Cartan, but were for a long time largely forgotten. The relevant geometry is a generalization of Felix Klein's celebrated {\it Erlanger Programm} \cite{Klein} to include non-homogeneous spaces, called  `Cartan geometries', or in Cartan's own terms, `{\it espaces g\'en\'eralis\'es}' \cite{Cartan2, Cartan3}.   The \MM\ gauge field is a special case of a `Cartan connection', which encodes geometric information relating the geometry of spacetime to the geometry of a homogeneous `model spacetime' such as de Sitter space.  Cartan connections have been largely supplanted in the literature by what is now the most familiar notion of `connection on a principal bundle' \cite{AMP}, first formalized by Cartan's student Charles Ehresmann \cite{Ehresmann}.   

While Ehresmann's connections---the sort that turned out to describe gauge fields of Yang--Mills theory---offer generality and flexibility not available in Cartan's version, some of Cartan's original geometric insights are lost in the abstraction.  
In this paper I review the essential ideas of Klein geometry, leading up to Cartan geometry. I show how Cartan geometry is ideally suited to describing the classical constraint problem of rolling a homogeneous manifold on another manifold, and use this idea to understand the geometry of \MM\ gravity.   I argue in particular for a return to Cartan's very `concrete' version of connections, as a means to better understanding of the geometry of gravity.

\subsubsection*{MacDowell--Mansouri gravity}

\MM\ gravity is based on a gauge theory with gauge group $G \supset \SO(3,1)$ depending on the sign of the cosmological constant:
\[
G= \left\{
\begin{array}{ll}
\SO(4,1) & \Lambda > 0 \\
\SO(3,2) & \Lambda < 0
\end{array}
\right.
\]
For sake of definiteness, let us focus on the physically favored $\Lambda>0$ case, where $G=\SO(4,1)$.  The key to the \MM\ approach is that the Lie algebra has a Killing-orthogonal splitting:
\begin{equation}
\label{so41split}
      \so(4,1)\iso \so(3,1)\oplus \R^{3,1},
\end{equation}
not as Lie algebras but as vector spaces, and this splitting is invariant under $\SO(3,1)$.  This lets us view the  Lorentz connection $\om$ and coframe field $e$ of Palatini-style \GR\ as two parts of a unified $\SO(4,1)$-connection $A$:
\[
         A =\om + \frac{1}{\ell} e
\]
where $\ell$ is a constant with units of length.   

This connection $A$ has a number of nice properties, as MacDowell and Mansouri noted.   Its curvature $F[A]$ also breaks up into $\so(3,1)$ and $\R^{3,1}$ parts.  The $\so(3,1)$ part is the curvature $R[\om]$  plus a cosmological constant term, while the $\R^{3,1}$ part is the torsion $d_\om e$:
\[
      F = \left(R  - \frac{\Lambda}{3}\,e\wedge e\right) + d_\om e 
\]
where we choose $\ell^2 = 3/\Lambda$.  
With this choice of scale, the curvature $F[A]$ vanishes precisely when $\om$ is the torson-free spin connection for a spacetime locally isometric to de Sitter space.

The gravity action used by MacDowell and Mansouri is:
\begin{equation}
\label{MM-action}
  S_{\rm \scriptscriptstyle MM}[A] = \frac{-3}{2G\Lambda} \int \tr(\widehat F \wedge \star \widehat F) .
\end{equation}
Here $\widehat F$ denotes the projection of $F$ into the subalgebra $\so(3,1)$, and $\star$ is an internal Hodge star operator.  The projection breaks the $\SO(4,1)$ symmetry down to $\SO(3,1)$.  Following MacDowell and Mansouri, we have broken symmetry in the Lagrangian by hand, but it is also possible to set the theory up for spontaneous symmetry breaking.  In any case, the resulting equations of motion are, rather surprisingly, Einstein's equations with cosmological constant $\Lambda$.  In fact, as we shall see, the \MM\ action and the usual Palatini action of general relativity are equivalent at the classical level, since they differ by a purely topological term:
 \[
     S_{\rm Pal} = S_{\rm\scriptscriptstyle MM} - \frac{3}{2G\Lambda}\int \tr R\wedge {\star R}. 
 \]

{\boldmath
\subsubsection*{The $BF$ reformulation}}

\noindent More recently, a different action for \MM\ gravity was proposed \cite{FreidelStarodubtsev,SmolinStarodubtsev,Staro}, in which the MacDowell--Mansouri connection $A$ is supplemented by an independent locally $\so(4,1)$-valued 2-form $B$: 
\beq
\label{FS-action}
       S =  \int \tr \left(B\wedge F - \frac{G\Lambda}{6}\widehat B \wedge  {\star \widehat B}\right).
\eeq
This action is equivalent to the original MacDowell--Mansouri action (\ref{MM-action}), by substituting the algebraic field equation  $B = \frac{3}{G\Lambda}\, F$ back into the action.  However, written in this new form,  \MM\ gravity has the appearance of a `deformation' of a topological gauge theory---the `$BF$ theory' \cite{Baez2} whose action is just the first term in (\ref{FS-action}). The symmetry breaking here occurs here only in the second term, with a dimensionless coefficient $G\Lambda \sim 10^{-120}$, suggesting that general relativity is in some sense `not too far' from a topological field theory.  

I explain both the original \MM\ action and the $BF$ reformulation in detail in Section~\ref{sec:MM}, after developing the appropriate geometric setting for such theories, which lies in Cartan geometry.   

\vspace{1em}

\subsubsection*{The idea of a Cartan geometry}

What is the geometric meaning of splitting an $\SO(4,1)$ connection into an $\SO(3,1)$ connection and coframe field?  For this it is easiest to first consider a lower-dimensional example, involving $\SO(3)$ and $\SO(2)$.  An oriented 2d Riemannian manifold is often thought of in terms of an $\SO(2)$ connection since, in the tangent bundle, parallel transport along two different paths from $x$ to $y$ gives results differing by a rotation of the tangent vector space at $y$:
\[
\xy
(2,-2)*{\includegraphics{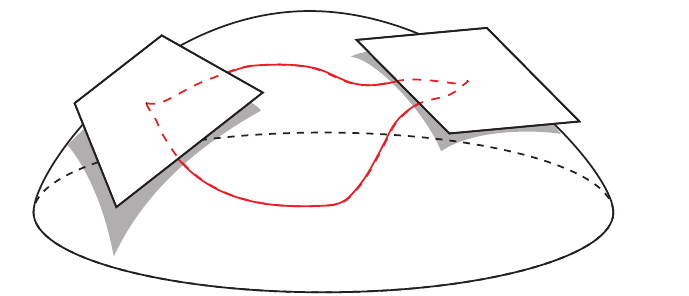}};
(28,-14)*{M};
(-19,3)*{x};
(-28,7)*{T_xM};
(16,5)*{y};
(28,7)*{T_yM};
\endxy
\]
In this context, we can ask the geometric meaning of extending the gauge group from $\SO(2)$ to $\SO(3)$.  The group $\SO(3)$ acts naturally not on the bundle $TM$ of tangent {\em vector spaces}, but on some bundle $SM$ of `tangent {\em spheres}'.  We can construct such a bundle, for example, by compactifying each fiber of $TM$.  Since $\SO(3)$ acts to rotate the sphere, an $\SO(3)$ connection on a 2d Riemannian manifold may be viewed as a rule for `parallel transport' of tangent spheres, which need not fix the point of contact with the surface:  
\[
\xy
(0,0)*{\includegraphics{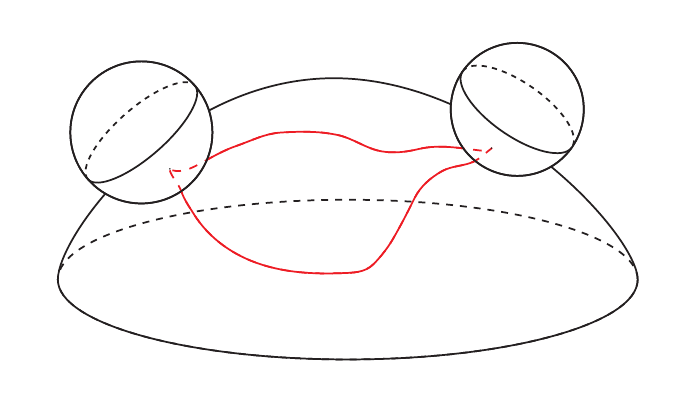}};
(28,-14)*{M};
(-19,3)*{x};
(-32,10)*{S_xM};
(16,5)*{y};
(29,7)*{S_yM};
\endxy
\]
An obvious way to get such an $\SO(3)$ connection is simply to {\em roll a ball on the surface}, without twisting or slipping.  Rolling a ball along two paths from $x$ to $y$ will in general give different results, but the results differ by an element of $\SO(3)$.  Such group elements encode geometric information about the surface itself.  

In our example, just as in the extension from the Lorentz group to the de Sitter group, we have an orthogonal splitting of the Lie algebra 
\[
  \so(3)\iso \so(2)\oplus \R^2
\]
given in terms of matrix components by
\[\footnotesize
   \left[ \begin{array}{ccc}
       0 & u & a  \\
       -u & 0 & b  \\
       -a & -b & 0  \\
  \end{array}\right] 
=
   \left[ \begin{array}{ccc}
       0 & u &   \\
       -u & 0 &   \\
        &  & 0  \\
  \end{array}\right] 
+   \left[ \begin{array}{ccc}
        &  & a  \\
        &  & b  \\
       -a & -b &   \\
  \end{array}\right].
\]
Like in the \MM\ case, this allows an $\SO(3)$ connection $A$ on an oriented 2d manifold to be split up into an $\SO(2)$ connection $\om$ and a coframe field $e$.  But here it is easy to see the geometric interpretation of these components: an infinitesimal rotation of the tangent sphere, as it begins to move along some path, breaks up into a part that rotates the sphere about its point of tangency and a part that moves the point of tangency:
\newsavebox{\rotation}
\savebox{\rotation}{
\begin{minipage}{4cm} \footnotesize  \raggedright 
The $\so(2)$ part gives an infinitesimal rotation around the axis through the point of tangency.
\end{minipage}}
\newsavebox{\transvection}
\savebox{\transvection}{
\begin{minipage}{4cm} \footnotesize  \raggedright The $\R^2$ part gives an infinitesimal translation of the point of tangency.
\end{minipage}}
\[
\xy
(0,0)*{\includegraphics{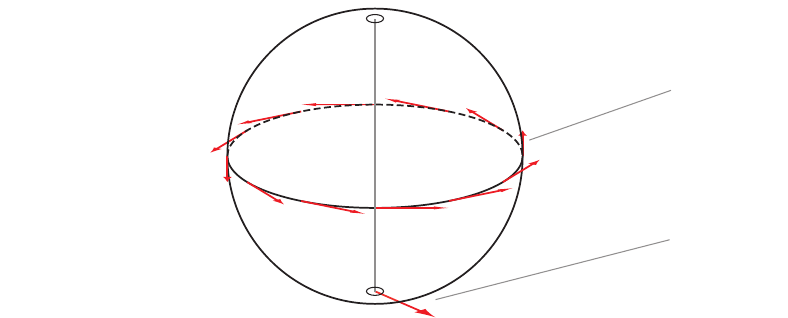}};
(51,12)*{\usebox{\rotation}};
(51,-10)*{\usebox{\transvection}};
\endxy
\]
The connection thus {\em defines} a method of rolling a sphere along a surface.

Extrapolating from this example to the extension $\SO(3,1)\subset \SO(4,1)$, we surmise a geometric interpretation for \MM\ gravity:  the $\SO(4,1)$ connection $A=(\om,e)$ encodes the geometry of spacetime $M$ by ``rolling de Sitter spacetime along $M$":
\[
\xy
(0,0)*{\includegraphics[height=5.5cm]{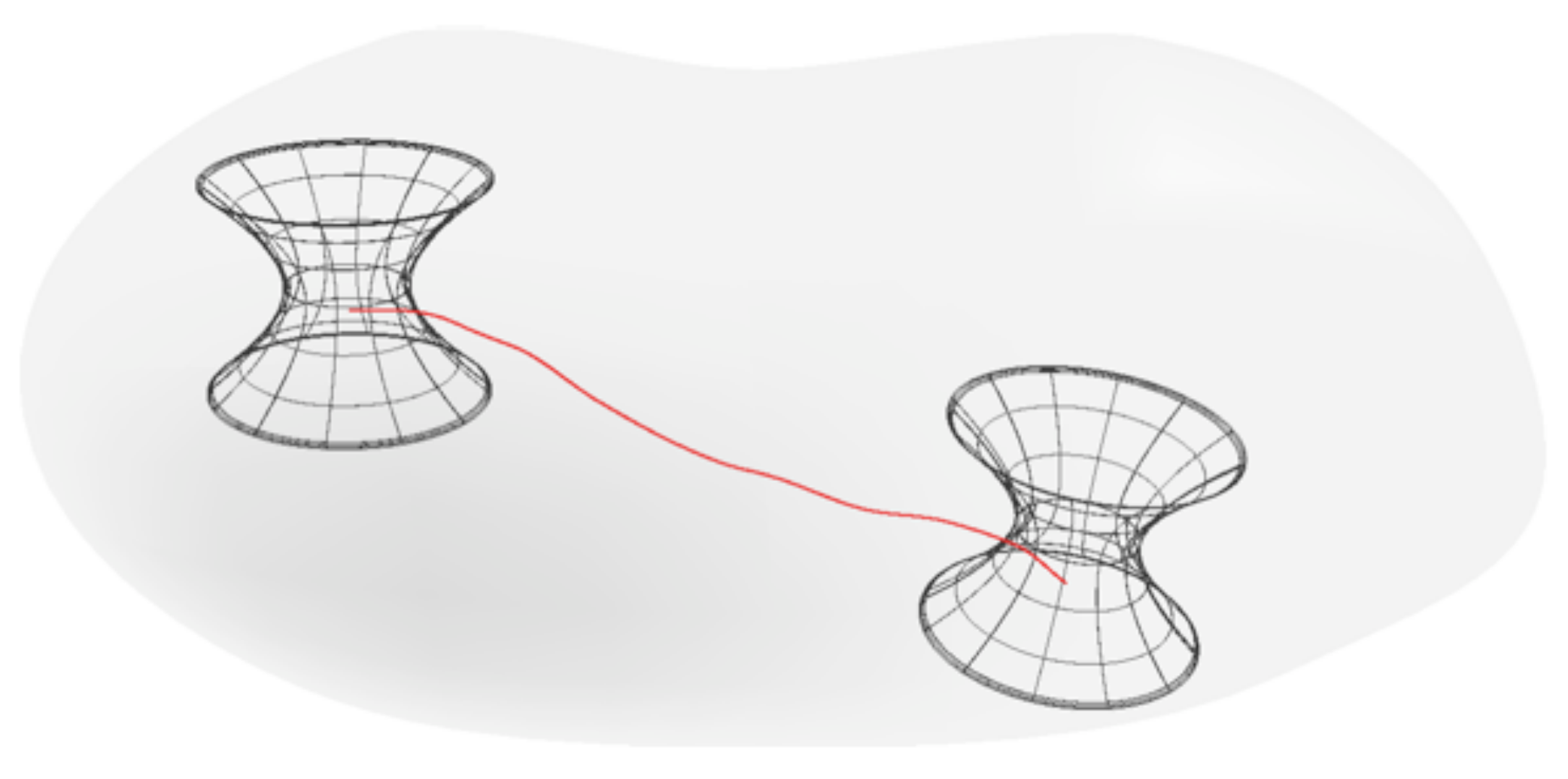}};
(-25,-10)*{\text{\begin{minipage}{4cm}\raggedright tangent de Sitter spacetime at $x\in M$\end{minipage}}};  
(25,7)*{\text{\begin{minipage}{4cm}\raggedright tangent de Sitter spacetime at $y\in M$\end{minipage}}};  
(46,-16)*{M};
\endxy
\]
This idea is appealing since, for spacetimes of positive cosmological constant, we expect de Sitter spacetime to be a better infinitesimal approximation than flat Minkowski vector space.  The geometric beauty of \MM\ gravity, and related approaches, is that they study spacetime using `tangent spaces' that are truer to the mean geometric properties of the spacetime itself.  Exploring the geometry of a surface $M$ by rolling a ball on it may not seem like a terribly useful thing to do if $M$ is a plane; if $M$ is some slight deformation of a sphere of the same radius, however, then exploring its geometry in this way is very sensible!   Likewise, approximating a spacetime by de Sitter space is most practical when the cosmological constant $\Lambda$ of spacetime equals the `internal' cosmological constant---the cosmological constant of the de Sitter model.

More generally, this idea of studying the geometry of a manifold by `rolling' another manifold---the `model geometry'---on it provides an intuitive picture of  `Cartan geometry'.   Cartan geometry, roughly speaking, is a generalization of Riemannian geometry obtained by replacing linear tangent spaces with more general homogeneous spaces.  As Sharpe explains in the preface to his textbook on the subject \cite{Sharpe}, Cartan geometry is the common generalization of Riemannian and Klein geometries.  Sharpe explains this neatly in a commutative diagram which we adapt here: 
\newsavebox{\EucGeom}
\savebox{\EucGeom}{
   \xy
      (0,2)*{\text{Euclidean}};
      (0,-2)*{\text{Geometry}};
    \endxy
}
\newsavebox{\KlGeom}
\savebox{\KlGeom}{
    \xy
      (0,2)*{\text{Klein}};
      (0,-2)*{\text{Geometry}};
    \endxy
}
\newsavebox{\CtnGeom}
\savebox{\CtnGeom}{
    \xy
      (0,2)*{\text{Cartan}};
      (0,-2)*{\text{Geometry}};
    \endxy
}
\newsavebox{\RiemGeom}
\savebox{\RiemGeom}{
    \xy
      (0,2)*{\text{Riemannian}};
      (0,-2)*{\text{Geometry}};
    \endxy
}
\newsavebox{\curvature}
\savebox{\curvature}{
    \xy 
       (0,2)*{\text{\footnotesize allow}};
       (0,0)*{\text{\footnotesize curvature}};
    \endxy
}
\newsavebox{\symmetry}
\savebox{\symmetry}{
    \xy 
       (0,2.5)*{\text{\footnotesize generalize}};
       (0,0)*{\text{\footnotesize symmetry group}};
    \endxy
}
\newsavebox{\tangspacegeom}
\savebox{\tangspacegeom}{
    \xy 
       (0,2.5)*{\text{\footnotesize generalize tangent}};
       (0,0)*{\text{\footnotesize space geometry}};
    \endxy
}
\[
\xygraph{
[]!{0;<5cm,0cm>:<0cm, 2cm>::}
[] {\usebox{\EucGeom}}="Euc" 
    :@{->}_{\usebox{\symmetry}} [d]
   {\usebox{\KlGeom}}="Klein" 
    :@{->}_{\usebox{\curvature}} [r]
   {\usebox{\CtnGeom}}="Ctn" 
"Euc" :@{->}^{\usebox{\curvature}} [r]  
    {\usebox{\RiemGeom}} :@{->}^{\usebox{\tangspacegeom}} "Ctn"
  }
\]
Like Euclidean geometry, a Klein geometry is homogeneous, meaning that there is a symmetry of the geometry taking any point to any other point.  Cartan geometry provides `curved' versions of arbitrary Klein geometries, in the same way that Riemannian geometry is a curved version of Euclidean geometry. 

But besides providing a beautiful geometric interpretation, and a global setting for the \MM\ way of doing gravity, Cartan geomety also helps in understanding the sense in which \MM\ theory is a deformation of a topological field theory.

\subsubsection*{Plan of the present paper}

In Section~\ref{sec:klein}, I briefly review Klein's viewpoint on homogeneous geometry using symmetry groups, focussing on the six examples most relevant to gravity: de Sitter, Minkowski, and anti de Sitter spacetimes, and their Wick-rotated versions, the spherical, Euclidean, and hyperbolic Riemannian spacetimes.  These six Klein geometries provide the homogeneous `model spacetimes' which are used to describe non-homogeneous spacetimes via Cartan geometry.  

Section~\ref{sec:cartan} provides an introduction to Cartan geometry, and explains why this is just the right sort of geometry to describe rolling a homogeneous space on a manifold.  In Section~\ref{sec:gauge}, I investigate further issues relevant to Cartan geometry, and particularly to the idea of doing gauge theory using a Cartan connection in place of the usual Ehresmann-type gauge field.

Section~\ref{sec:pal-to-mm} focusses on viewing general relativity through the lens of Cartan geometry.  I begin with a review of the Palatini formalism  
and show how it can be viewed in terms of Cartan geometry.   
This leads naturally to the construction of the \MM\ action from the Palatini action; I discuss the \MM\ action and its $BF$ reformulation.

\section{Homogeneous spacetimes and Klein geometry}
\label{sec:klein}

Klein revolutionized modern geometry with the realization that almost everything about a homogeneous geometry---with a very broad interpretation of what constitutes a `ge\-om\-e\-try'---is encoded in its groups of symmetries.   From the Kleinian perspective, the objects of study in geometry are `homogeneous spaces'.  While many readers will be familiar with homogeneous geometry, the idea is essential to understanding Cartan geometry, so I review it here in some detail. 

\subsection{Klein geometry}

A {\bf homogeneous space} $(G,X)$ is an abstract space $X$ together with a group $G$ of transformations of $X$, such that $G$ acts transitively: given any $x,y\in X$ there is some $g\in G$ such that $gx =y$. There is a deliberate ambiguity here about what sort of `abstract space' $X$ should be.  In different applications, $X$ might be a mere discrete set, a topological space, a Riemannian manifold, etc.  For our immediate purposes, the most important cases are when $X$ has at least the structure of a smooth manifold, and $G$ acts as diffeomorphisms. 

The main tools for exploring a homogeneous space $(G,X)$ are subgroups $H\subgp G$ which preserve, or `stabilze', interesting `features' of the geometry.  What constitutes an interesting feature of course depends on the geometry.  For example, Euclidean geometry, $(\R^n,\ISO(n))$, has points, lines, planes, polyhedra, and so on, and one can study subgroups of the Euclidean group $\ISO(n)$ which preserve any of these.  `Features' in other homogeneous spaces may be thought of as generalizations of these notions.  We can also work backwards, {\em defining} a feature of a geometry abstractly as that which is preserved by a given subgroup.   If $H$ is the subgroup preserving a given feature, then the space of all such features of $X$ may be identified with the coset space $G/H$:
\[
   G/H = \{ gH : g\in G\} = \text{ the space of ``features of type $H$"}.
\]

Let us illustrate why this is true using the most basic of features, the feature of `points'.  Given a point $x\in X$, the subgroup of all symmetries $g\in G$ which fix $x$ is called the {\bf stabilizer}, or {\bf isotropy group} of $x$, and will be denoted $H_x$.  Fixing $x$, the transitivity of the $G$-action implies we can identify each $y\in X$ with the set of all $g\in G$ such that $gx = y$.  If we have two such symmetries:
\[
     gx = y \qquad g'x = y
\]
then clearly $g^{-1}g'$ stabilizes $x$, so $g^{-1}g' \in H_x$.  Conversely, if $g^{-1}g' \in H_x$ and $g$ sends $x$ to $y$, then $g'x=gg^{-1}g'x = gx = y$.  Thus, the two symmetries move $x$ to the same point if and only if $gH_x = g'H_x$.  The points of $X$ are thus in one-to-one correspondence with cosets of $H_x$ in $G$.  Better yet, the map $f\maps X\to G/H_x$ induced by this correspondence is $G$-equivariant:
\[
       f(gy) = gf(y) \qquad \forall g \in G, y \in X
\]
so $X$ and $G/H_x$ are isomorphic as $H$-spaces.

All this depends on the choice of $x$, but if $x'$ is another point, the stabilizers are conjugate subgroups: 
\[
     H_x=gH_{x'} g^{-1}
\]
where $g\in G$ is any element such that $gx' = x$.  Since these conjugate subgroups of $G$ are all isomorphic, it is common to simply speak of ``the" point stabilizer $H$, even though fixing a particular one of these conjugate subgroups gives implicit significance to the points of $X$ fixed by $H$.   By the same looseness of vocabulary, the term `homogeneous space' often refers to the coset space $G/H$ itself.

To see the power of the Kleinian point of view, consider a familiar example of a homogeneous space: $(n+1)$-dimensional Minkowski spacetime.  While this is most obviously thought of as the `space of events', there are other interesting `features' to Minkowski spacetime, and the corresponding homogeneous spaces each tell us something about the geometry of special relativity.  The group of symmetries preserving orientation and time orientation is the connected Poincar\'e group $\ISO_0(n,1)$.  The stabilizer of an event is the connected Lorentz group $\SO_0(n,1)$ consisting of boosts and rotations.  The stabilizer of an event {\em and} a velocity is the group of spatial rotations around the event, $\SO(n)$.  The stabilizer of a spacelike hyperplane is the group of Euclidean transformations of space, $\ISO(n)$.  This gives us a piece of the lattice of subgroups of the Poincar\'e group, with corresponding homogeneous spaces:
\newsavebox{\events}
\savebox{\events}{\xy (0,0)*{\ISO_0(n,1)/\SO_0(n,1)};
    (0,-5)*{\text{\footnotesize `event space' (Minkowski)}};
    \endxy}
\newsavebox{\velocity}
\savebox{\velocity}{\xy (0,0)*{\SO_0(n,1)/\SO(n)};
    (0,-5)*{\text{\footnotesize `velocity space' (hyperbolic)}};
    \endxy}
\newsavebox{\spc}
\savebox{\spc}{\xy (0,0)*{\ISO(n)/\SO(n)};
    (0,-5)*{\text{\footnotesize `position space' (Euclidean)}};
    \endxy}
\newsavebox{\frames}
\savebox{\frames}{\xy (0,0)*{\ISO_0(n,1)/\ISO(n)};
    (0,-5)*{\text{\footnotesize `space of spacelike hyperplanes'}};
    \endxy}
\[
\xygraph{
  []!{0;<2.4cm,-2.5cm>:<2.5cm, 2.5cm>::}
  []{\ISO_0(n,1)}="A" :@{-}^{\usebox{\events}} [r] 
    {\SO_0(n,1)} :@{-}^{\usebox{\velocity}} [d]
    {\SO(n)} :@{-}^{\usebox{\spc}} [l]
    {\ISO(n)} :@{-}^{\usebox{\frames}} "A"
  }
\]

`Klein geometries', for the purposes of this paper, will be certain types of homogeneous spaces.  The geometries we are interested in are all `smooth' geometries, so we require that the symmetry group $G$ be a Lie group.  We also require the subgroup $H$ to be a closed subgroup of $G$.  This is obviously necessary if we want the quotient $G/H$ to have a topology where 1-point subsets are closed sets.  In fact, the condition that $H$ be closed in $G$ suffices to guarantee $H$ is a Lie subgroup and $G/H$ is a smooth homogeneous manifold.  

We also want Klein geometries to be connected.  Leaving this requirement out is sometimes useful, particularly in describing discrete geometries.  However, our purpose is not Klein geometry \textit{per se}, but Cartan geometry, where the key idea is comparing a manifold to a `tangent Klein geometry'.  Connected components not containing the `point of tangency' have no bearing on the Cartan geometry, so it is best to simply exclude disconnected homogeneous spaces from our definition.

\begin{defn}
\label{Klein-geom}
A (smooth, connected) {\bf Klein geometry} $(G,H)$ consists of a Lie group $G$ with closed subgroup $H$, such that the coset space $G/H$ is connected. 
\end{defn}

As Sharpe emphasizes \cite{Sharpe}, for the purposes of understanding Cartan geometry it is useful to view a Klein geometry $(G,H)$ as the principal right $H$-bundle 
\[
\xygraph{ 
  [] {G} :@{->}  [d] 
 {G/H}
      }
\]
This is a principal bundle since the fibers are simply the left cosets of $H$ by elements of $G$, and these cosets are isomorphic to $H$ as right $H$-sets.  

Strictly speaking, a `homogeneous space' clearly should not have a preferred basepoint, whereas the identity coset  $H\in G/H$ is special.   It would thus be better to define a Klein geometry to be a principal $H$-bundle $P\to X$ which is merely {\em isomorphic to} the principal bundle $G \to G/H$:
\[
\xygraph{ 
  [] {P}="P" :@{->}_{\pi} [d]
    {X} :@{->}^{\sim} [r] 
    {G/H}="gh"
  "P" :@{->}^{\sim} [r] 
  {G} :@{->} "gh" 
    }
\]
but not canonically so.  For our purposes, however, it will actually be convenient to have an obvious basepoint in the Klein geometry.  Since we are interested in approximating the local geometry of a manifold by placing a Klein geometry {\em tangent} to it, the preferred basepoint $H\in G/H$ will serve naturally as the `point of tangency'.    

\subsection{Metric Klein geometry}

For studying the essentially distinct types of Klein geometry, it is enough to consider the coset spaces $G/H$.  However, for many applications, including \MM, one is interested not just in the symmetry properties of the homogeneous space, but also in its metrical properties.  If we wish to distinguish between spheres of different sizes, or de Sitter spacetimes of different cosmological constants, for example, then we need more information than the symmetry groups.  For such considerations, we make use of the fact that there is a canonical isomorphism of vector bundles \cite{Sharpe}
\[
\xygraph{ 
  %[]!{0;<4cm,0cm>:}
  []!{0;<2.4cm,0cm>:<1.2cm, 1.8cm>::}
  []{T(G/H)}="TM" :@{->}^{\sim} [r] {G\times_H \g/\h}
                        :@{->}^{\pi} [d] {G/H}="M"
        "TM"        :@{->}_{p} "M" 
  }
\] 
where the bundle on the right is the bundle associated to the principal bundle $G\to G/H$ via the adjoint representation on $\g/\h$. This means the space of tangent vectors at any point in the Klein geometry $G/H$ may be identified with $\g/\h$, and an $\Ad(H)$-invariant metric on $\g/\h$ induces a {\em homogeneous} metric on the tangent bundle $T(G/H)$.  In physically interesting examples, this metric will generally be nondegenerate of Riemannian or Lorentzian signature.  One way to obtain such a metric is to use the Killing form on $\g$, which is invariant under $\Ad(G)$, hence under $\Ad(H)$, and passes to a metric on $\g/\h$.  When $\g$ is semisimple the Killing form is nondegenerate.  But even when $\g$ is not semisimple, it may be possible to find a nondegenerate $H$-invariant metric on $\g/\h$, hence on $T(G/H)$.  This leads us to define:

\begin{defn}
A {\bf metric Klein geometry} $(G,H,\eta)$ is a Klein geometry $(G,H)$ equipped with a (possibly degenerate) $\Ad(H)$-invariant metric $\eta$ on $\g/\h$.  
\end{defn}

Notice that any Klein geometry can be made into a metric Klein geometry in a trivial way by setting $\eta=0$.  In cases of physical interest, it is usually possible to choose $\eta$ to be nondegenerate. 

\subsection{Homogeneous model spacetimes}
\label{sec:modelspacetimes}
For \MM\ gravity, there are 4 homogeneous spacetimes we are most interested in, corresponding to Lorentzian or Riemannian gravity with cosmological constant either positive or negative.  These are the de Sitter, anti de Sitter, spherical, and hyperbolic models.  We can also consider the $\Lambda \to 0$ limits of these, the Minkowski and Euclidean models.  This gives us six homogeneous `model spacetimes', each of which can be described as a Klein geometry $G/H$: 
\[
\begin{array}{lccc}
   & \Lambda<0 &   \Lambda = 0 & \Lambda>0 \\ 
\text{Lorentzian}\rule{0cm}{.8cm}
   &  \begin{array}{c}\text{\footnotesize \sf anti de Sitter} \\ \SO(3,2)/\SO(3,1) \end{array} 
   &  \begin{array}{c}\text{\footnotesize \sf Minkowski} \\ \ISO(3,1)/\SO(3,1)  \end{array}  
   &  \begin{array}{c}\text{\footnotesize \sf de Sitter} \\  \SO(4,1)/\SO(3,1)  \end{array} \\
\text{Riemannian}\rule{0cm}{.8cm} 
   & \begin{array}{c}\text{\footnotesize \sf hyperbolic} \\ \SO(4,1)/\SO(4)  \end{array}
   & \begin{array}{c}\text{\footnotesize \sf Euclidean} \\ \ISO(4)/\SO(4)  \end{array}
   & \begin{array}{c}\text{\footnotesize \sf spherical} \\ \SO(5)/\SO(4)  \end{array}
\end{array}
\]

For many purposes, these spacetimes can be dealt with simultaneously, with the cosmological constant as a parameter.  Let us focus on the three Lorentzian cases in what follows; their Riemannian counterparts can be handled in the same way.  In their fundamental representations, the Lie algebras $\so(4,1)$, $\Iso(3,1)$ and $\so(3,2)$ consist of $5\times 5$ matrices of the form:
\[\footnotesize
   \left[ \begin{array}{rrrrr}
       0^{\phantom 1} & b^1 & b^2 & b^3 & p^0/\ell \\
       b^1 & 0^{\phantom 1} & j^3 & -j^2 & p^1/\ell \\
       b^2 & -j^3 & 0^{\phantom 1} & j^1 & p^2/\ell \\
       b^3 & j^2 & -j^1 & 0^{\phantom 1} & p^3/\ell \\
       \xe p^0/\ell & -\xe p^1/\ell & -\xe p^2/\ell & -\xe p^3/\ell  & 0^{\phantom 1} \\
  \end{array}\right] 
    =
    j^i J_i + b^i B_i +  \frac{1}{\ell}p^a P_a.
\]
Here $J_i$, $B_i$ are generators of rotations and boosts, $P_a = (P_0,P_i)$ are generators of translations,  $\xe$ is the sign of the cosmological constant:
\beq
\label{sign}
    \xe = \left\{ 
      \begin{array}{rl}
         1\rule{1em}{0pt} & \g = \phantom{\frak i}\so(4,1) \\
         0 \rule{1em}{0pt}& \g = \Iso(3,1) \\
         -1\rule{1em}{0pt} & \g =  \phantom{\frak i}\so(3,2)\, ,
      \end{array}
      \right. 
\eeq
and we have introduced a length scale $\ell$ so that $p^a$ may be identified with the components of a `translation vector' on the homogeneous spacetime.  
Commutation relations are of course dependent on $\xe$: 
\[
\begin{array}{lll}
[J_i,J_j] = -\varepsilon_{ijk}J^k &  &  \\   
{[B_i,J_j] = \varepsilon_{ijk}B^k} & [B_i,B_j] = \varepsilon_{ijk}J^k & \\ 
{[P_i,J_j] = -\varepsilon_{ijk}P^k} & [P_i,B_j] = -P_0\xd_{ij} & {[P_0,J_i] = 0}\\
{ [P_i,P_j] = -\xe\, \varepsilon_{ijk}J^k} & [P_0, P_i] = -\xe\, B_i & [P_0, B_i] = -P_i 
\end{array}
\]

All of these model spacetimes are naturally nondegenerate {\em metric} Klein geometries.   For the cases with $\xe\ne 0$, we can equip the Lie algebra $\g$ with a nondegenerate invariant metric:
\beq
\label{killing}
      \langle \xi, \zeta \rangle = - \frac{\xe}{2} \tr(\xi \zeta)
\eeq
which is proportional to the Killing form.  With respect to this metric, we have the orthogonal, $\Ad(\SO(3,1))$-invariant direct sum decomposition:
\[
      \g = \so(3,1) \oplus \p 
\] 
where the subalgebra $\so(3,1)$ is spanned by the rotation and boost generators $J_i,B_i$, and the complement $\p\iso\g/\so(3,1)\iso \R^{3,1}$ is spanned by the $P_a$.  The restriction of the metric to $\p$ is the Minkowski metric with signature $({-}{+}{+}{+})$, and we obtain a metric Klein geometry, as defined in the previous section, simply by translating this metric around the homogeneous space.

The choice of scale $\ell$ (and $\xe=\pm 1$) effectively selects the value of the cosmological constant to be
\beq
\label{internal}
          \Lambda = \frac{3\xe}{\ell^2}
\eeq 
To see this, let us take a closer look at the de Sitter case, where $\Lambda>0$.  De Sitter spacetime is most easily pictured as the 4-dimensional submanifold of 5d Minkowski space given by
\[
   \MdS  = \left\{(t,w,x,y,z)\in \R^{4,1}\; \left|\; -t^2+w^2+x^2+y^2+z^2 = \frac{3}{\Lambda} \right.\right\} 
\]
The group $G=\SO(4,1)$ acts in the usual way on the ambient $(4+1)$-dimensional space, and the subgroup $H\iso\SO(3,1)$ in the upper $4\times 4$ block is the stabilizer of the point 
$$x_o = (0,\ldots,0,\sqrt{3/\Lambda}).$$  
The intention of introducing the length scale $\ell$ is that the element $\frac{1}{\ell} p^\mu P_\mu \in \p$ should be identified with $p^\mu\partial_\mu\in T_{x_o}\MdS$ in the Klein geometry, via the exponential map.  That is,
\[
    \frac{d}{ds} \exp\left({ \frac{s}{\ell}}\,p^\mu P_\mu\right)x_o \Bigg|_{t=0} = 
     { \frac{1}{\ell} \sqrt{\frac{3}{\Lambda}}}\,p^\mu \partial_\mu 
\]
should equal $p^\mu\partial_\mu$, and hence we should take
\[
   \Lambda=\frac{3}{ \ell^2}.
\]
The expression $\eta_{ab}v^a w^b$ may then be interpreted either as the metric (\ref{killing}) applied to $v,w\in \p$ or as the metric of de Sitter space applied to the counterparts of $v,w$ tangent to $\MdS$ at $x_o$.  The argument for the $\Lambda< 0$ case is the same except for a sign, and in either case we obtain the claimed value (\ref{internal}) for the cosmological constant.

When the cosmological constant vanishes, the situation is a bit more subtle.  The isometry group of Minkowski space, $\ISO(3,1)$, does not have a nondegenerate adjoint-invariant metric on its Lie algebra.   In fact, the metric induced by the trace {\em vanishes} on the subspace corresponding to $\R^{3,1}$.  However, we require a metric on this subspace to be invariant only under $\SO(3,1)$, not under the full Poincar\'e group.  Such a metric is easily obtained, noting the semidirect product structure:
\[
        \Iso(3,1) = \so(3,1) \ltimes \R^{3,1}
\]  
of the Poincar\'e Lie algebra.  Using the trace on $\so(3,1)$ together with the usual Minkowski metric on $\R^{3,1}$ gives an nondegenerate $\SO(3,1)$-invariant metric on the entire Poincar\'e Lie algebra.  In particular, the metric on the $\R^{3,1}$ part makes $\ISO(3,1)/\SO(2)$ into a nondegenerate metric Klein geometry.  

Notice that in the Minkowski case, as in de Sitter or anti de Sitter, identifying the `translation' subspace of $\Iso(3,1)$ with spacetime tangent vectors still involves choosing a length $\ell$ by which to scale vectors.  But now this choice is not constrained by the value of the cosmological constant.  This points out a key difference beween the $\Lambda = 0$ and $\Lambda \neq 0$ cases:  Minkowski spacetime has an extra `rescaling' symmetry that is broken as the cosmological constant becomes nonzero. 

When $M$ is one of our homogeneous model spacetimes, one can, of course, calculate the Riemann curvature for the metric $g_{\mu\nu}$ on $TM$ induced by the metric $\eta_{ab}$ on $\g/\h$.  The result is:
\[
 R_{\mu\nu\rho\sigma} 
     = \frac{\Lambda}{3}\left(g_{\mu\rho}g_{\nu\sigma} - g_{\mu\sigma}g_{\nu\rho}\right)
\]
Equivalently, if $e^a_\mu$ is a coframe field, locally identifying each $T_xM$ with $\g/\h\iso \R^{3,1}$, our homogeneous spacetimes satisfy:
\[
 {R^{ab}}_{\mu\nu} 
     = \frac{\Lambda}{3}\left(e^a_{\mu}e^b_{\nu} - e^a_{\nu}e^b_{\mu}\right)
\]
In this paper, I use form notation rather than spacetime indices; my conventions are given by:
\[
    R^{ab} = \half {R^{ab}}_{\mu\nu}\, dx^\mu\wedge dx^\nu = \frac{\Lambda}{3} (e^a_\mu \, dx^\mu) \wedge (e^b_\nu \, dx^\nu) = \frac{\Lambda}{3}\; e^a \wedge e^b.
\]
I often suppress internal indices as well, so the local condition for spacetime to be homogeneous with cosmological constant $\Lambda$ may thus be written simply
\begin{equation}
\label{model.curvature}
     R = \frac{\Lambda}{3}\; e \wedge e.
\end{equation}

\section{Cartan geometry}
\label{sec:cartan}
While the beauty of Klein's perspective on geometry is widely recognized, the spacetime we live in is clearly not homogeneous.  This does not imply, however, that Kleinian geometry offers no insight into actual spacetime geometry!  Cartan discovered a beautiful generalization of Klein geometry---a way of modeling non-homogeneous spaces as `infinitesimally Kleinian'.   The goal of this section is to explain this idea as it relates to spacetime geometry.    

While this section and the next are intended to provide a fairly self-contained introduction to basic Cartan geometry, I refer the reader to the references for further details on this very rich subject.  In particular, the articles by Alekseevsky and Michor \cite{AlekMic} and Ruh \cite{Ruh} and the book by Sharpe \cite{Sharpe} are helpful resourses, and serve as the main references for my explanation here.

\subsection{Ehresmann connections}
Before giving the definition of Cartan connection, I review the more familiar notion of an Ehresmann  connection on a principal bundle.  Such a connection is just the type that shows up in ordinary gauge theories, such as Yang--Mills.  My purpose in reviewing this definition is merely to easily contrast it with the definition of a `Cartan connection', to be given in Section~\ref{sec:cartan-geom}.

Both Ehresmann and Cartan connections are related---in somewhat different ways---to the Maurer--Cartan form, the canonical 1-form any Lie group $G$ has, with values in its Lie algebra $\g$:
\[
           \omega_G \in \Omega^1 ( G , \frakg ).
\]
This 1-form is simply the derivative of left multiplication in $G$:
\begin{align*}
        \omega_G &\maps T G \to \frakg \\
               \omega_G &(x) :=   (L_{g^{-1}})_\ast(x)  \quad \forall x\in T_g G. 
\end{align*}
Since the fibers of a principal $G$-bundle look just like $G$, they inherit a Maurer--Cartan form in a natural way.   Explicitly, the action of $G$ on a principal right $G$-bundle $P$ is such that, if $P_x$ is any fiber and $y\in P_x$,  the map
\begin{align*}
   G &\to P_x \\
  g&\mapsto yg
 \end{align*}
is invertible.  The inverse map lets us pull the Maurer--Cartan form back to $P_x$ in a unique way:
\[
   T_{xg}P_x \to T_g G \to \g
\]   
Because of this canonical construction, the 1-form thus obtained on $P$ is also called a {\bf Maurer--Cartan form}, and denoted $\omega_G$.  

Ehresmann connections can be defined in a number of equivalent ways \cite{AMP}.  The definition we shall use is the following one.  
\begin{defn}
\label{ehresmann}
An {\bf Ehresmann connection} on a principal right $H$-bundle 
\[
\xygraph{ 
  [] {P} :@{->}^{\pi} [d] 
      {M}  }
\]
is an $\frakh$-valued 1-form $\omega$ on $P$
\[
        \omega\maps TP \to \frakh
\]
satisfying the following two properties:
\begin{enumerate}
  \item $R_h^\ast \omega = \Ad(h^{-1})\omega$ for all $h\in H$;
  \item $\omega$ restricts to the Maurer--Cartan form $\omega_H\maps TP_x \to \h$ on fibers of $P$.
\end{enumerate}
\end{defn}
Here $R^\ast_h \om$ denotes the pullback of $\om$ by the right action
\begin{align*}
     R_h \maps P &\to P \\
      p &\mapsto ph 
\end{align*}
of $h\in H$ on $P$.

The {\bf curvature} of an Ehresmann connection $\om$ is given by the familiar formula
\[
       \Omega[\om] = d\omega + \half [\omega, \omega]
\] 
where the bracket of $\h$-valued forms is defined using the Lie bracket on Lie algebra parts and the wedge product on form parts. 

\subsection{Definition of Cartan geometry} 
\label{sec:cartan-geom}

We are ready to state the formal definition of Cartan geometry, essentially as given by Sharpe \cite{Sharpe}.

\begin{defn}
\label{def:cartan}
A {\bf Cartan geometry} $({\pi\maps P\to M},A)$ modeled on the Klein Geometry $(G,H)$ is a principal right $H$-bundle 
\[
\xygraph{ 
  [] {P} :@{->}^{\pi} [d] 
      {M}  }
\]
equipped with a $\frakg$-valued 1-form $A$ on $P$
\[
        A\maps TP \to \frakg
\]
called the {\bf Cartan connection}, satisfying three properties:
\begin{enumerate}
\item[0.] For each $p\in P$, $A_p\maps T_p P\to \frakg$ is a linear isomorphism;
\item $(R_h)^\ast A = \Ad(h^{-1})A \quad \forall h \in H$;
\item $A$ takes values in the subalgebra $\h \subseteq \g$ on vertical vectors, and in fact restricts to the Maurer--Cartan form $\omega_H\maps TP_x \to \h$ on fibers of $P$.
\end{enumerate}
\end{defn}

Compare this definition to the definition of Ehresmann connection.  The most obvious difference is that the Cartan connection on $P$ takes values not in the Lie algebra $\h$ of the gauge group of the bundle, but in the larger algebra $\g$.  The addition of the $0$th requirement in the above definition has important consequences.  Most obviously, $G$ must be chosen to have the same dimension as $T_p P$.  In other words, the Klein geometry $G/H$ must have the same dimension as $M$.  In this way Cartan connections have a more ``concrete" relationship to the base manifold than Ehresmann connections, which have no such dimensional restrictions.  Also, the isomorphisms $A\maps T_p P \to \g$ may be inverted at each point to give an injection
\[
     X_A\maps \g \to \Vect(P)
\]   
so any element of $\g$ gives a vector field on $P$.  The restriction of $X_A$ to the subalgebra $\h$ gives vertical vector fields on $P$, while the restriction of $X_A$ to a complement of $\h$ gives vector fields on the base manifold $M$ \cite{AlekMic}. 

Cartan geometries also inherit any additional structures on the tangent spaces of their model Klein geometries.  In particular, when $G/H$ is a metric Klein geometry, i.e.\ when it is equipped with an $H$-invariant metric on $\g/\h$, $M$ inherits a metric of the same signature, via the isomorphism $T_xM \iso \g/\h$, which comes from the isomorphism $T_pP \iso \g$.

The {\bf curvature} of a Cartan connection is given by the same formula as in the Ehresmann case:
\[
       F[A] = dA + \half[A,A].
\]
This curvature is a 2-form valued in the Lie algebra $\g$.  It can be composed with the canonical projection onto $\g/\h$:
\[
\xymatrix{
     \Lambda^2(TP) \ar[r]^{\quad\; F} \ar@/_4ex/[rr]_{T} & \g \ar[r] & \g/\h
}
\]
and the composite $T$ is called the {\bf torsion}; as explained in Section~\ref{sec:reductive}, it is the natural generalization of the sort of `torsion' familiar from ordinary Riemannian geometry. 

The simplest examples of Cartan geometries are Klein geometries.  Indeed, if $G$ is a Lie group, then its Maurer--Cartan form $\om_G$ is a canonical Cartan connection for the Klein geometry $G\to G/H$, 
for {\em any} closed subgroup $H\subseteq G$. The well-known `structural equation' for the Maurer--Cartan form, 
\beq
\label{structuralequation}
     d\om_G = - \half [\om_G,\om_G],
\eeq
is interpreted in this context as the statement of vanishing Cartan curvature.

\subsection{Geometric interpretation: rolling Klein geometries}
\label{sec:hamster}

In Section~\ref{sec:intro}, I claimed that Cartan geometry is about ``rolling the model Klein geometry on the manifold."  Let us now see why a Cartan geometry on $M$ modeled on $G/H$ contains just the right data to describe the idea of rolling $G/H$ on $M$.   To understand this, we return to the example of the sphere rolling on a surface $M$ embedded in $\R^3$.   For this example we have
\begin{align*}
       G&=\SO(3)\\
       H&=\SO(2)
\end{align*}
and the model space is $S^2 = \SO(3)/\SO(2)$. The Cartan geometry consists of a principal $\SO(2)$-bundle $P$ over $M$ together with a 1-form $\omega\in \Omega^1(M,\so(3))$ satisfying the three properties above. 

To understand the geometry, it is helpful to consider the situation from the point of view of an `observer' situated at the point of tangency between the `real' space and the homogeneous model.  In fact, in the rolling ball example, such an observer is easily imagined.  Picture the model sphere as a ``hamster ball''---a type of transparent plastic ball designed to put a hamster or other pet rodent in to let it run around the house without getting lost.  But here, the hamster gets to run around on some more interesting, more lumpy surface than your living room floor, such as a Riemann surface:
\[
\xy
(0,0)*{\includegraphics[height=5cm]{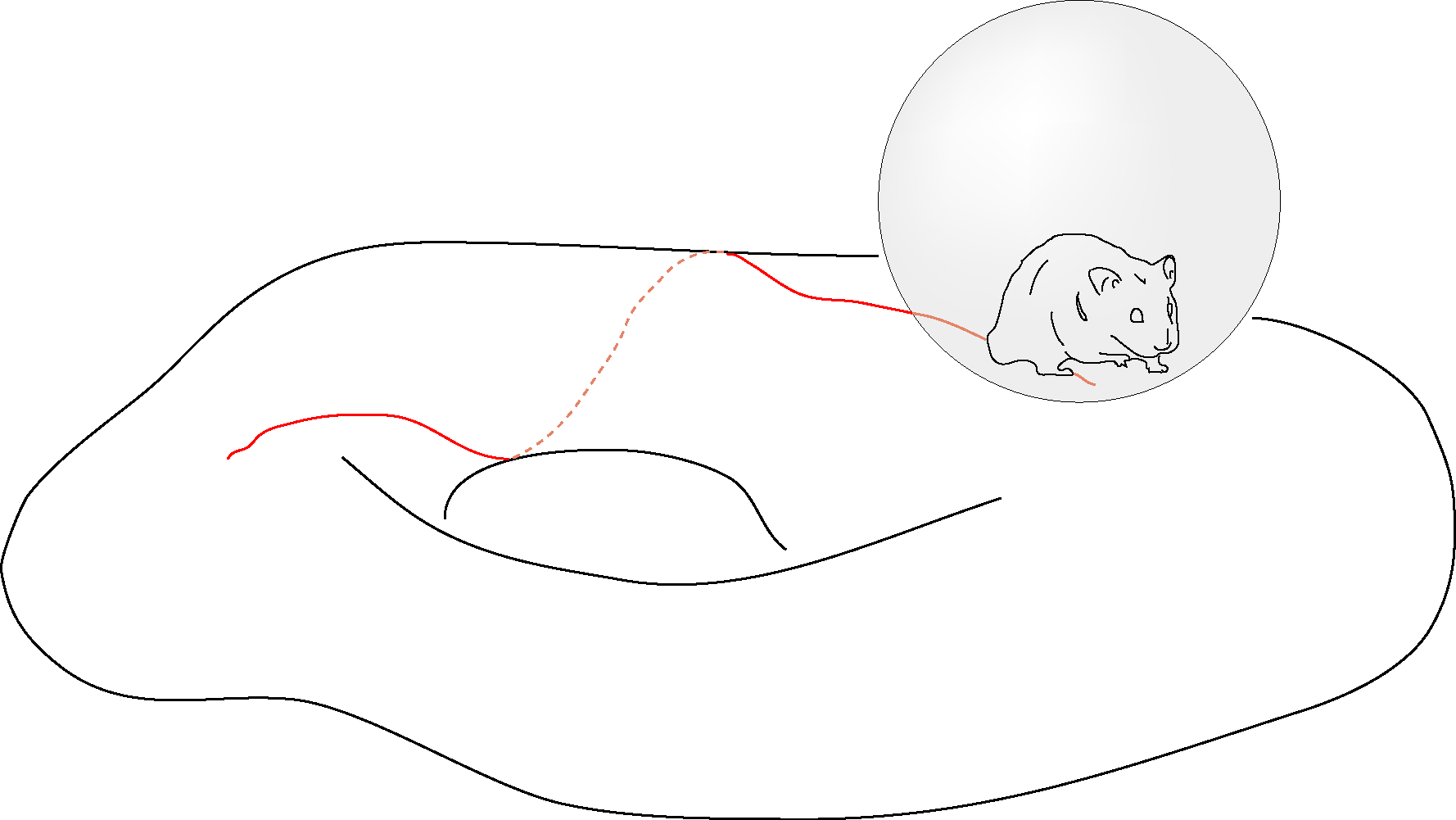}}
\endxy
\]
It may sound silly, but this is the easiest way to begin to visualize Cartan connections!  
In this context, what is the geometric meaning of the $\SO(2)$-bundle $P$ in the definition of Cartan geometry?  Essentially, $P$ can be thought of as the bundle of `hamster configurations', where a hamster configuration is specified by the hamster's position on the surface $M$, together with the direction the hamster is facing.  

One key point, which is rather surprising on first sight, is that our bundle $P$ tells us nothing about the configuration of the rolling sphere itself.  It tells us only where the hamster is, and which direction he is pointing.  Naively, we might try describing the rolling of a ball on a surface using the space of all configurations of the ball itself, which would be a principal $\SO(3)$-bundle over the surface.  But in fact, our principal $\SO(2)$-bundle is sufficient to describe rolling {\em without slipping or twisting}.  This becomes obvious when we consider that the motion of the hamster completely determines the motion of the ball. 

Now a Cartan connection:
\[
     A\maps TP \to \so(3)
\]
takes `infinitesimal changes in hamster configuration' and gives infinitesimal rotations of the sphere he is sitting inside of.  An `infinitesimal change in hamster configuration' consists of a tiny rotation together with a `transvection'--- a pure translation of the point of tangency.  The resulting element of $\so(3)$ is the tiny rotation of the sphere, as seen by the hamster.

I now describe in detail the geometric interpretation of conditions 0, 1, and 2 in the definition of a Cartan connection, in the context of this example. 
\begin{enumerate}
\item[0.] {\em $A_p\maps T_p P \to \so(3)$ is a linear isomorphism.}   The hamster can move in such a way as to produce any tiny rotation of the sphere desired, and he can do this in just one way.  In the case of a tiny rotation that lives in the stabilizer subalgebra $\so(2)$, note that the sphere's rotation is always viewed relative to the hamster: his corresponding movement is just a tiny rotation of his body, while fixing the point of tangency to the surface.  In particular, the isomorphism is just the right thing to impose a `no twisting' constraint.   Similarly, since the hamster can produce any transvection in a unique way, the isomorphism perfectly captures the idea of a `no slipping' constraint. 

\item[1.] {\em $(R_h)^\ast A = \Ad(h^{-1})A$ for all $h \in \SO(2)$.}  This condition is `$\SO(2)$-equivariance', and may be interpreted as saying there is no absolute significance to the specific direction the hamster is pointing in.  A hamster rotated by $h \in \SO(2)$ will get different elements of $\so(3)$ for the same infinitesimal motion, but they will differ from the elements obtained by the unrotated hamster by the adjoint action of $h^{-1}$ on $\so(3)$.   

\item[2.] {\em $A$ restricts to the $\SO(2)$ Maurer-Cartan form on vertical vectors.}  A vertical vector amounts to a slight rotation of the hamster inside the hamster ball, without moving the point of tangency.  Using the orientation, there is a canonical way to think of a slight rotation of the hamster as an element of $\so(2)$, and $A$ assigns to such a motion precisely this element of $\so(2)$.

\end{enumerate}

Using this geometric interpretation, it is easy to see that the model Klein geometries themselves serve as the prototypical examples of flat Cartan geometries.  Rolling a Klein geometry on {\em itself} amounts to simply moving the point of tangency around.  Thus, just as $\R^n$ has a canonical way of identifying all of its linear tangent spaces, $S^n$ has a canonical way of identifying all of its tangent spheres, $\Hyp^n$ has a canonical way of identifying all of its tangent hyperbolic spaces, and so on.  

It is perhaps worth mentioning another example---an example that is sort of `dual' to the hamster ball rolling on a flat plane---which I find equally instructive.  Rather than a hamster in a sphere, exploring the geometry of a plane, consider a person (a  15th century European, say) standing on a flat planar `model Earth', tangent to the actual, spherical Earth.  The plane rolls as she steps, the point of tangency staying directly beneath her feet.  This rolling gives an $\ISO(2)/\SO(2)$ Cartan geometry on the Earth's surface.  She can even use the rolling motion to try drawing a local map of the Earth on the plane.  As long as she doesn't continue too far, the rolling is slight, so the map will be fairly accurate.

In \MM\ gravity, we are in a related geometric situation.  The principal $\SO(3,1)$-bundle describes possible event/velocity pairs for an ``observer".  This observer may try drawing a map of spacetime $M$ by rolling Minkowski spacetime along $M$, giving an $\ISO(3,1)/\SO(3,1)$ Cartan connection.  A smarter observer, if $M$ has $\Lambda>0$, might prefer to get an $\SO(4,1)/\SO(3,1)$ Cartan connection by rolling a model of de Sitter spacetime along $M$.   

\subsection{Reductive Cartan geometry}
\label{sec:reductive}

The most important special case of Cartan geometry for our purposes is the `reductive' case.  Since $\h$ is a vector subspace of $\g$, we can always write 
\[
  \frakg \iso \frakh \oplus \frakg/\frakh
\] 
as vector spaces.  A Cartan geometry is said to be {\bf reductive} if the quotient $\g/\h$ may be identified with an  $\Ad(H)$-invariant subspace of $\g$.  In other words, when the geometry is reductive, the above direct sum is a direct sum of $\Ad(H)$-representations.   
A reductive Cartan connection $A$ may thus be written as 
\[
     A= \omega + e   \qquad 
     \begin{array}{l}
        \omega \in \Omega^1(P,\h)\\
        e \in \Omega^1(P,\g/\h)
    \end{array}
\]
Diagrammatically:
\[
\objectmargin={3pt}
\xygraph{ 
  []!{<0cm,0cm>;<1cm,-1cm>:<1cm, 1cm>::}
 [] {TP}="TP"  [r] [u]  {\frakg}="g"
        "TP"        :@{->}^{A} "g" 
                :@{->} [u] {\frakh}="h"
            "g" :@{->} [r] {\frakg/\frakh}="gmodh"
        "TP" :@/^{.35cm}/^{\omega} "h"
        "TP" :@/_{.35cm}/_{e} "gmodh"
  }
\]
It is easy to see that the $\h$-valued form $\omega$ is simply an Ehresmann connection on $P$, and we interpret the $\g/\h$-valued form $e$ as a generalized {\bf coframe field}.   

The concept of a reductive Cartan connection provides a geometric foundation for the \MM\ action.  In particular, it gives global meaning to the trick of combining the local connection and coframe field 1-forms of general relativity into a connection valued in a larger Lie algebra.  Physically, for theories like \MM, the reductive case is most important because gauge transformations of the principal $H$-bundle act on $\g$-valued forms via the adjoint action.  The $\Ad(H)$-invariance of the decomposition says gauge transformations do not mix up the `connection' parts with the `coframe' parts of a reductive Cartan connection.

One can of course use the $\Ad(H)$-invariant decomposition of $\g$ to split any other $\g$-valued differential form into $\h$ and $\g/\h$ parts.  Most importantly, we can split the curvature $F$ of the Cartan connection $A$:
\[
\objectmargin={3pt}
\xygraph{ 
  []!{<0cm,0cm>;<1cm,-1cm>:<1cm, 1cm>::}
 [] {\Lambda^2(TP)}="TP2"  [r] [u]  {\frakg}="g"
        "TP2"        :@{->}^{\qquad F} "g" 
                :@{->} [u] {\frakh}="h"
            "g" :@{->} [r] {\frakg/\frakh}="gmodh"
        "TP2" :@/^{.35cm}/^{\widehat F} "h"
        "TP2" :@/_{.35cm}/_{T} "gmodh"
  }
\]
The $\g/\h$ part $T$ is the {\bf torsion}.  The $\h$ part $\widehat F$ is related to the curvature of the Ehresmann connection $\omega$, but there is an important difference: {\em The `curvature' $\widehat F$ is the Ehresmann curvature modified in such a way that the model Klein geometry becomes the standard for `flatness'.}   In other words, $\widehat F$ vanishes when the geometry is locally that of $G/H$.  

To understand this claim, consider the most basic example of a `flat' Cartan geometry: the Klein geometry $G/H$ itself, whose canonical Cartan connection is the Maurer--Cartan form $\om_G$.   When the geometry is reductive, we can write $\omega_G = \om_H + e$, where $\om_H$ is the Maurer--Cartan form on the fibers of $G\to G/H$.  By the structural equation (\ref{structuralequation}), $G/H$ is `flat' in the Cartanian sense.  In particular, this means both $\h$ and $\g/\h$ parts of the curvature must vanish, even though the geometry certainly needn't be `flat' in the more traditional Riemannian sense of the word. 

To make these ideas more concrete, let us work out the components of the curvature in the cases most relevant to gravity.  The six Klein model spacetimes listed in Section~\ref{sec:klein}---de Sitter, Minkowski, anti de Sitter, and their Riemannian analogs---are all reductive.  (In fact, they are `symmetric spaces'; this is why the curvature formulas below particularly simple \cite{symmetric}.) For any of these models, the Cartan connection is an $\g$-valued 1-form $A$ on a principal $H$-bundle, which we take to be the frame bundle $FM$ on spacetime:
\[
       A \in \Omega^1(FM, \g).
\]
We identify $\g/\h$ with Minkowski space $\R^{3,1}$ (or Euclidean $\R^4$ in the Riemannian cases) by picking a unit of length $\ell$.

In index notation,  we write the two parts of the connection as
\[
      {A^a}_b =  {\om^a}_b \quad \text{and} \quad  {A^a}_4 =\frac{1}{\ell}e^a.
\]
This implies
\[
         {A^4}_b = \frac{-\xe}{\ell}\, e_b,
\]
where $\xe$ is chosen according to the choice of $\g$, by (\ref{sign}). We use these components to calculate the two parts of the curvature 
\[
{F^I}_J= d{A^I}_J + {A^I}_K \wedge {A^K}_J
\]
as follows.  For the  $\so(3,1)$ part, we get:
\begin{align*}
     {F^a}_b&= {dA^a}_b + {A^a}_c \wedge {A^c}_b + {A^a}_4 \wedge {A^4}_b \\
               &=  {d\om^a}_b + {\om^a}_c \wedge {\om^c}_b   
                                           - \frac{\xe}{\ell^2}\, e^a \wedge e_b \\
               &= {R^a}_b  - \frac{\xe}{\ell^2}\, e^a \wedge e_b
\end{align*}
where $R$ is the curvature of the $\SO(3,1)$ Ehresmann connection $\om$; for the $\R^{3,1}$ part:
\begin{align*}
    {F^a}_4 &= {dA^a}_4 + {A^a}_b \wedge {A^b}_4 \\
                           &= \frac{1}{\ell} \left( de^a + {\om^a}_b\wedge e^b \right) \\
                           &= \frac{1}{\ell}\; d_\om e^a.
\end{align*}
Hence
\beq
\label{reductive.curvature}
         F = \Big(R - \frac{\xe}{\ell^2}e\wedge e\Big) + \frac{1}{\ell} d_\om e.
\eeq
The same calculations hold formally in the Riemannian analogs as well, the only difference being that indices are lowered with $\delta_{ij}$ rather than $\eta_{ij}$. 

We now see clearly, by (\ref{reductive.curvature}), what it means for a Cartan connection $A=\om + e$ based on any of our six models to be flat:
\[
       F = 0 \qquad \iff \qquad R =\frac{\xe}{\ell^2} e \wedge e   \; \text{ and } \;  d_\om e = 0
\]
Comparing to (\ref{model.curvature}), these are precisely the local equations characterizing the torsion-free spin connection for a homogeneous spacetime, provided we take the cosmological constant 
\[
          \Lambda =  \frac{3\xe}{\ell^2}\, .
\]
Naturally, this is the cosmological constant (\ref{internal}) of the model itself. 

Borrowing language from Yang--Mills theory, it is helpful to think of the cosmological constant (\ref{internal}) of the model homogeneous spacetime as a sort of `internal cosmological constant'.  The criterion for a spacetime to be flat in the Cartanian sense is then that spacetime have purely cosmological curvature where the spacetime cosmological constant matches the internal one.  
One {\em could} try describing spacetime with cosmological constant $\Lambda$ using a model whose cosmological constant is $\lambda \neq \Lambda$, but this is not a very natural thing to do.   In fact, though, this unnatural description is essentially what is used in the standard approach to general relativity:  ordinary semi-Riemannian geometry---that is, $\lambda =0$ Cartan geometry---is used to describe spacetimes with $\Lambda\neq 0$.

If we agree to use a model spacetime with cosmological constant $\Lambda$,  the parts of a reductive connection and its curvature can be summarized diagrammatically in the three Lorentzian cases as follows:
\[
\objectmargin={3pt}
\xygraph{ 
  []!{<0cm,0cm>;<1cm,-1cm>:<1cm, 1cm>::}
 [] {T(FM)}="TP"  [r] [u]  {\g}="g"
        "TP"        :@{->}^{A} "g" 
                :@{->} [u] {\so(3,1)}="h"
            "g" :@{->} [r] {\R^{3,1}}="gmodh"
        "TP" :@/^{.35cm}/^{\om} "h"
        "TP" :@/_{.35cm}/_{\frac{1}{\ell}\,e} "gmodh"
  }
\objectmargin={3pt}
\xygraph{ 
  []!{<0cm,0cm>;<1.5cm,-1cm>:<1.5cm, 1cm>::}
 [] {\Lambda^2(T(FM))}="TP2"  [r] [u]  {\g}="g"
        "TP2"        :@{->}^{\qquad F} "g" 
                :@{->} [u] {\so(3,1)}="h"
            "g" :@{->} [r] {\R^{3,1}}="gmodh"
        "TP2" :@/^{.35cm}/^{R-\frac{\Lambda}{3}\, e\wedge e} "h"
        "TP2" :@/_{.35cm}/_{\frac{1}{\ell}\;d_\om e} "gmodh"
  }
\]
where $\ell$ and $\Lambda$ are related by the equation 
\[
        \ell^2\Lambda = 3\xe.
\]
As observed earlier, for $\Lambda=\xe=0$ the value of $\ell^2$ is not constrained by the cosmological constant, so there is an additional scaling symmetry in Cartan geometry modelled on Minkowski or Euclidean spacetime.

As a final note on reductive Cartan geometries, in terms of the constituent fields $\om$ and $e$, the {\bf Bianchi identity} 
\[
    d_A F = 0
\]
for a reductive Cartan connection $A$ breaks up into two parts.  One can show that these two parts are the Bianchi identity for $\om$ and another familiar identity:
\[
     d_\om R = 0 \qquad d_\om^2 e = R\wedge e.
\]

\section{Cartan-type gauge theory}
\label{sec:gauge}

Part of the case I wish to make is that gravity---particularly in \MM-like formulations---should be seen as based on a type of gauge theory where the connection is not an Ehresmann connection but a Cartan connection.  Unlike gauge fields in `Ehresmann-type' gauge theories, like Yang-Mills theory, the gravitational field does not carry purely `internal' degrees of freedom.  Cartan connections give a concrete correspondence between spacetime and a Kleinian model, in a way that is ideally suited to a geometric theory like gravity. 

In this section, I discuss issues---such as holonomy and parallel transport---relevant to doing gauge theory with a Cartan connection as the gauge field.  Some of these issues are clarified by considering certain associated bundles of the Cartan geometry.

\subsection{A sequence of bundles}
\label{sec:exact}

Just as Klein geometry involves a sequence of $H$-spaces: 
\[
       H \to G \to G/H,
\]
Cartan geometry can be seen as involving the induced sequence of bundles:
\newsavebox{\hbdle}
\savebox{\hbdle}{
   \xy
     (0,2)*{ \text{\footnotesize principal}};
     (0,-2)*{\text{\footnotesize $H$-bundle}}
   \endxy
}
\newsavebox{\gbdle}
\savebox{\gbdle}{
   \xy
     (0,2)*{ \text{\footnotesize principal}};
     (0,-2)*{\text{\footnotesize $G$-bundle}}
   \endxy
}
\newsavebox{\tangklein}
\savebox{\tangklein}{
   \xy
     (0,2)*{ \text{\footnotesize bundle of tangent}};
     (0,-2)*{\text{\footnotesize Klein geometries}}
   \endxy
}
\[
\xygraph{ 
  %[]!{0;<4cm,0cm>:}
  []!{0;<3cm,0cm>:<0cm, 3.4cm>::}
  []{P}="1" :@{->}^{\displaystyle \iota} [r] {P\times_H G}="2" :@{->} [r] {P\times_H G/H}="3"
                 "1"  :@{->}_{\usebox{\hbdle}} [dr] M="M"
                "2"  :@{->}|-{\usebox{\gbdle}} "M"
                "3"  :@{->}^{\usebox{\tangklein}} "M" 
  }
\]
The bundle $Q = P\times_H G\to M$ is associated to the principal $H$-bundle $P$ via the action of $H$ by left multiplication on $G$.  This $Q$ is a principal right $G$-bundle, and the map
\begin{align*}
     \iota\maps P &\to P\times_H G \\
            p &\mapsto [p,1_G]
\end{align*}
is a canonical inclusion of $H$-bundles.  I call the associated bundle $\kappa\maps P\times_H G/H\to M$, the {\bf bundle of tangent Klein geometries}.  This is an appropriate name, since it describes a bundle over $M$ whose fibers are copies of the Klein geometry $G/H$, each with a natural `point tangency'.  Explicitly,  for $x\in M$, the {\bfm Klein geometry tangent to $M$ at $x$} is the fiber $\kappa^{-1} x$, and the {\bf point of tangency} in this tangent geometry is the equivalence class $[p,H]$ where $p$ is any point in $P_x$ and $H$ is the coset of the identity.  This is well defined since any other `point of tangency' is of the form $[ph,H] = [p,hH] = [p,H]$, where $h\in H$.

There is an important relationship between Cartan connections on $P$ and Ehresmann connections on $Q = P\times_H G$.  First, given a Cartan connection $A\maps TP\to \g$, one can use equivariance to extend to an Ehresmann connection $\tilde A\maps TQ \to \g$ in a unique way, so that $\iota^\ast{\tilde A} = A$.  We call $\tilde A$ the {\bf associated Ehresmann connection} for the Cartan connection $A$.  Not every Ehresmann connection is associated in this way to some Cartan connection on $P$, but precisely those Ehresmann connections $\tilde A\maps TQ \to \g$ for which $\ker \, {\tilde A} \cap \iota_\ast(TP) =0$.  Thus, we may think of a Cartan connection as an Ehresmann connection on $Q$ satisfying this additional property \cite{Sharpe}.  

From another perspective, one can define a notion of {\bf generalized Cartan connection}, in which the 0th requirement in Definition~\ref{def:cartan}, that $A_p\maps T_pP \to \g$ be an isomorphism, is replaced by the weaker requirement that $T_p$ and $\g$ have the same dimension.  This amounts to allowing the coframe field to be degenerate.  Then if 
$\tilde A\maps TQ \to \g$  is an arbitrary Ehresmann connection on $Q$, $A:= \iota^\ast{\tilde A}\maps TP \to \g$ is a generalized Cartan connection on $P$.  
So, generalized Cartan connections on $P$ are in one-to-one correspondence with Ehresmann connections on $Q$ \cite{AlekMic}. 

%There is an interesting correspondence between Cartan connections on $P$ and Ehresmann connections on $Q = P\times_H G$.  To understand this correspondence, we introduce the notion of a {\bf generalized Cartan connection} \cite{AlekMic}, in which we replace the 0th requirement in Definition~\ref{def:cartan}, that $A_p\maps T_pP \to \g$ be an isomorphism, by the weaker requirement that $T_p$ and $\g$ have the same dimension.   It is not hard to show that if 
%$$\tilde A\maps TQ \to \g$$ 
%is an Ehresmann connection on $Q$ then 
%\[
%             A:= \iota^\ast{\tilde A}\maps TP \to \g
%\]
%is a generalized Cartan connection on $P$.  In fact, given a generalized Cartan connection on $P$, there is a unique Ehresmann connection $\tilde A$ on $Q$ such that $A= \iota^\ast{\tilde A}$, so the generalized Cartan connections on $P$ are in one-to-one correspondence with Ehresmann connections on $Q$ \cite{AlekMic}.   Moreover, the generalized Cartan connection $A$ associated to an Ehresmann connection $\tilde A$ on $Q$ is a Cartan connection if and only if $\ker \tilde A \cap \iota_\ast(TP) =0$ \cite{Sharpe}.  

\subsection{Parallel transport in Cartan geometry} 

In a spacetime of positive cosmological constant, how does one decide how much the geometry deviates from that of de Sitter spacetime?  From the Cartan perspective, one way is to do parallel transport in the bundle of tangent de Sitter spacetimes.

There are actually two things we might mean by `parallel transport' in Cartan geometry.  First, if the geometry is reductive, then the $\h$ part of the $G/H$-Cartan connection is an Ehresmann connection $\om$.  We can use this Ehresmann connection to do parallel transport in the bundle of tangent Klein geometries in the usual way.  Namely, if 
\[
  \xg \maps [t_0,t_1] \to M
\]
is a path in the base manifold, and $[p,gH]$ is a point in the tangent Klein geometry at $\xg(t_0)$, then the translation of $[p,gH]$ along $\xg$ is
\[
      [\tilde\xg(t),gH]
\] 
where $\tilde \xg$ is the horizontal lift of $\xg$ starting at $p\in P$.  However, this method, aside from being particular to the reductive case, is also not the sort of parallel transport that is obtained by rolling the model geometry, as in our intuitive picture of Cartan geometry.  In particular, the translation of the point of tangency $[p,H]$ of the tangent Klein geometry at $x=\xg(t_0) \in M$ is always just the point of tangency in the tangent Klein geometry at $\xg(t)$.  This is expected, since the gauge group $H$ only acts in ways that stabilize the basepoint.  We would like to describe a sort of parallel transport that does not necessarily fix the point of tangency. 

The more natural notion of parallel transport in Cartan geometry does not require the geometry to be reductive.  A Cartan connection cannot be used in the same way as an Ehresmann connection to do parallel transport, because Cartan connections do not give `horizontal lifts'.   Horizontal subspaces are given by the kernel of an Ehresmann connection; Cartan connections have no kernel.  To describe the general notion of parallel transport in a Cartan geometry, we make use of the associated Ehresmann connection on $Q= P\times_H G$, as described in the previous section. 

To understand general parallel transport in Cartan geometry, observe first that we have a canonical isomorphism of fiber bundles
\[
\xygraph{ 
  %[]!{0;<4cm,0cm>:}
  []!{0;<3cm,0cm>:<1.5cm, 2cm>::}
  []{P\times_H G/H}="1" :@{->}^{\displaystyle \iso} [r] {Q\times_G G/H}="2" 
                "1"  :@{->} [dr] M="M"
                "2"  :@{->} "M"
  }
\] 
To see this, note first that the $H$-bundle inclusion map $\iota\maps P \to Q$ induces an inclusion of the associated bundles by
\begin{align*}
     \iota' \maps P\times_H G/H &\to Q \times_G G/H \\
       [p,gH] &\mapsto [\iota(p),gH].
\end{align*}
This bundle map has an inverse which we construct as follows.  An element of $Q\times_G G/H = P \times_H G \times_G G/H$ is an equivalence class $[p,g',gH]$, with $p\in P$, $g' \in G$, and $gH\in G/H$.  Any such element can be written as $[p,1,g'gH]$, so we can define a map that simply drops this ``1" in the middle:
\begin{align*}
     \phi\maps Q\times_G G/H &\to P\times_H G/H \\
       [p,g',gH] &\mapsto [p,g'gH].
\end{align*}
It is easy to check that this is a well-defined bundle map, and 
\begin{gather*}
     \phi\iota' [p,gH] = \phi[\iota(p),gH] = \phi[p,1,gH] = [p,gH] \\
     \iota'\phi[p,g',gH] = \iota'[p,g'gH] = [p,1,g'gH] = [p,g',gH]
\end{gather*}
so $\iota'=\phi^{-1}$ is a bundle isomorphism.

While these are isomorphic as fiber bundles, the isomorphism is not an isomorphism of associated bundles (in the sense described by Isham \cite{Isham}), since it does not come from an isomorphism of the underlying principal bundles.  In fact, while $P\times_H G/H$ and $Q\times_G G/H$ are isomorphic as fiber bundles, there is a subtle difference between the two descriptions: the latter bundle does not naively have a natural `point of tangency' in each fiber, except via the isomorphism $\iota'$.  Indeed, the choice of a point of tangency in each fiber of $Q\times_G G/H\iso Q/H$ is precisely the trivializing secion that reduces $Q$ to $P$. 

Given the above isomorphism of fiber bundles, and given the associated Ehresmann connection defined in the previous section, we have a clear prescription for parallel transport.  Namely, given any $[p,gH]$ in the tangent Klein geometry at $x = \xg(t_0)\in M$, we think of this point as a point in $Q\times_G G/H$, via the isomorphism $\iota'$, use the Ehresmann connection on $Q$ to translate along $\xg$, then turn the result back into a point in the bundle of tangent Klein geometries, $P \times_H G/H$, using $\phi$.  That is, the parallel transport is 
\[
                  \phi([ {\widehat \xg}(t), gH])
\] 
where
\[
      \widehat \xg\maps [t_0,t_1] \to Q
\]
is the horizontal lift of $\xg\maps [t_0,t_1] \to M$ starting at $\iota(p)\in Q$, with respect to the Ehresmann connection associated with the Cartan connection on $P$.  Note that this sort of parallel transport need not fix the point of tangency.

\subsection*{Holonomy and development}

Just as a Cartan geometry has two notions of parallel translation, it also has two notions of holonomy, taking values in either $G$ or $H$.  Whenever the geometry is reductive, we can take the holonomy along a loop using the Ehresmann connection part of the Cartan connection.  This gives a holonomy for each loop with values in $H$.  In fact, without the assumption of reductiveness, there is a general notion of this $H$ holonomy, which I shall not describe.  In general there is a topological obstruction to defining this  type of holonomy of a Cartan connection: it is not defined for all loops in the base manifold, but only those loops that are the images of loops in the principal $H$-bundle \cite{Sharpe}. 

The other notion of holonomy, with values in $G$, can of course can be calculated by relying on the associated Ehresmann connection on $Q= P\times_H G$.

Besides holonomies around loops, a Cartan connection gives a notion of `development on the model Klein geometry'.  Suppose we have a Cartan connection $A$ on $P\to M$ and a piecewise-smooth path in $P$, 
\[
      \xg \maps [t_0,t_1] \to P,
\]
lifting a chosen path in $M$.  Given any element $g\in G$, the {\bfm development of $\xg$ on $G$} starting at $g$ is the unique path 
\[
     \xg_G \maps [t_0,t_1] \to G
\]
such that $\xg_G (t_0) = g$ and  $\xg^\ast A = \xg_G^\ast \om_G \in \Omega^1([t_0,t_1],\g)$.  To actually calculate the development, one can use the usual path-ordered exponential
\[
     \xg_G (t) =  Pe^{-\int_0^t \om(\tilde \xg'(s)) ds} \in G.
\]

Composing $\xg_G$ with the quotient map $G \to G/H$ gives a path on the model Klein geometry:
\[
      \xg_{G/H} \maps [t_0,t_1] \to G/H.
\]
called the {\bfm development of $\xg$ on $G/H$} starting at $gH$.  This path is independent of the lifting $\xg$, depending only on the path in the base manifold $M$.  \cite{Sharpe}
 
 In the $\SO(3)/\SO(2)$ `hamster ball' example of Section~\ref{sec:hamster}, the development is the path traced out on the ball itself by the point of tangency on the surface, as the ball rolls. 
 
{\boldmath
\subsection{$BF$ Theory and flat Cartan connections}
\label{sec:BF}
}
\noindent As an example of a gauge theory with Cartan connection, let us consider using a Cartan connection in the topological gauge theory known as `$BF$ theory' \cite{Baez2}.  Special cases of such a theory have already been considered by Freidel and Starodubtsev, in connection with \MM\ gravity \cite{FreidelStarodubtsev}, but without the explicit Cartan-geometric framework.  

In ordinary $BF$ theory with gauge group $H$, on $n$-dimensional spacetime, the fields are an Ehresmann connection $A$ on a principal $H$-bundle $P$, and an $\Ad(P)$-valued $(n-2)$-form $B$, where 
\[
         \Ad(P) = P \times_H \h
\]
is the vector bundle associated to $P$ via the adjoint representation of $H$ on its Lie algebra.  Denoting the curvature of $A$ by $F$, the $BF$ theory action
\[
       S_{BF} = \int \tr(B\wedge F)
\]
leads to the equations of motion:
\begin{align*}
   F&= 0\\
   d_A B &= 0.
\end{align*}
That is, the connection $A$ is flat, and the field $B$ is covariantly closed.

We wish to copy this picture as much as possible using a Cartan connection of type $G/H$ in place of the Ehresmann $H$-connection.   Doing so requires, first of all, picking a Klein model $G/H$ of the same dimension as $M$.  For the $B$ field, the obvious analog is an $(n-2)$-form with values in the bundle
\[
        \Ad_\g(P) := P \times_H \g
\]
where $H$ acts on $\g$ via the restriction of the adjoint representation of $G$.  Formally, we obtain the same equations of motion
\begin{align*}
   F&= 0\\
   d_A B &= 0.
\end{align*}
but these must now be interpreted in the Cartan-geometric context.  

In particular, the equation $F=0$ says the Cartan connection is flat.  In other words, `rolling' the tangent Klein geometry on spacetime is trivial, giving an isometric identification between any contractible neighborhood in spacetime and a neighborhood of the model geometry $G/H$.  Of course, the rolling can still give nontrivial holonomy around noncontractible loops.  This indicates that solutions of Cartan-type $BF$ theory are related to `geometric structures' \cite{Thurston}, which have been used to study a particular low-dimensional case of $BF$ theory, namely, 3d quantum gravity \cite{Carlip2}. 

Let us work out a more explicit example: Cartan-type $BF$ theory based on one of the $(3+1)$-dimensional reductive models discussed in Sections \ref{sec:modelspacetimes} and \ref{sec:reductive}.  Each of these geometries is reductive, so we can decompose our $\g$-valued fields $A$, $F$, and $B$ into $\so(3,1)$ and $\R^{3,1}$ parts.  We have done this for $A$ and $F$ already, in Section \ref{sec:reductive}, where we had:
\[
      {A^a}_b =  {\om^a}_b \;, \qquad  {A^a}_4 =\frac{1}{\ell}e^a,
\]
\[
     {F^a}_b = {R^a}_b  - \frac{\xe}{\ell^2}\, e^a \wedge e_b\;, \qquad 
     {F^a}_4 = \frac{1}{\ell}\; d_\om e^a.
\]
For $B$, let $\bs=\widehat B$ denote the $\so(3,1)$ part, and $\bt$ the $\R^{3,1}$ part, so 
\[
 {B^a}_b= {\bs^a}_b \qquad {B^a}_4 = \frac{1}{\ell}\bt^a.
\]
Note that this gives
\[
        {B^4}_b = - \frac{\xe}{\ell} \bt_b\;. 
\]
with $\xe$ the sign of the cosmological, as in (\ref{sign}).
We also need to know how to write $d_A B$ in terms of these component fields.  We know that
\begin{align*}
    {d_A B}^{IJ} :&= d B^{IJ} + [A,B]^{IJ} \\
               &= dB^{IJ} + {A^I}_K \wedge B^{KJ} - {B^I}_K \wedge A^{KJ} ,
\end{align*}
so for both indices between 0 and 3 we have
\begin{align*}
    d_A B^{ab} :&= d B^{ab} +  {A^a}_c \wedge B^{cb} - {B^a}_c \wedge A^{cb} 
                                                  +  {A^a}_4\wedge B^{4b} - {B^a}_4\wedge A^{4b} \\
                                 &= d_\om \bs^{ab} - \frac{\xe}{\ell} e^{a} \wedge \bt^b + 
                                 \frac{\xe}{\ell} \bt^{a} \wedge e^b
\end{align*}
and for an index 4,
\begin{align*}
 d_A B^{a 4} = d_A \bs^{a 4} &= d B^{a 4} + {A^a}_b \wedge B^{b 4} - {B^a}_b \wedge A^{b 4} \\
              & = d_\om \bt^a - \frac{1}{\ell}{\bs^a}_b \wedge e^{b}.
\end{align*}
The equations for $BF$ theory with Cartan connection based on de Sitter, Minkowski, or anti de Sitter model geometry are thus
\begin{align*}
   R - \frac{\xe}{\ell^2} e \wedge e &= 0\\
   d_\om e &= 0 \\
      d_\om  \bs + \frac{\xe}{\ell^2}(\bt\wedge e - e \wedge \bt) &= 0 \\
      d_\om \bt - \frac{1}{\ell} \bs\wedge e &= 0 
\end{align*}
In terms of the constituent fields of the reductive geometry, classical Cartan-type $BF$ theory is thus described by the Levi-Civita connection on a spacetime of purely cosmological curvature, with constant $\Lambda = 3\xe/\ell^2$, together with an pair of auxiliary fields $\bs$ and $\bt$, satisfying two equations.  We shall encounter equations very similar to these in the $BF$ reformulation of MacDowell--Mansouri gravity. 

\newpage

\section{From Palatini to \MM}
\label{sec:pal-to-mm}

In this section, I show how thinking of the standard Palatini formulation of general relativity in terms of  Cartan geometry leads in a natural way to the \MM\ formulation.     

\subsection{The Palatini formalism}
\label{sec:palatini}

Before describing the \MM\ approach, I briefly recall in this section the better-known Palatini formalism.  The main purpose in doing this is to firmly establish the global differential geometric setting of the Palatini approach, in order to compare to that of \MM.  Experts may safely skip ahead after skimming to fix notation.

In its modern form, the Palatini formalism downplays the metric $g$ on spacetime, which plays a subordinate role to the {\bf coframe field} $e$, a vector bundle morphism:
\[
\xygraph{ 
  %[]!{0;<4cm,0cm>:}
  []!{0;<1.5cm,0cm>:<.75cm, 1.8cm>::}
  []{TM}="TM" :@{->}^{e} [r] {\fake}
                        :@{->}^{\pi} [d] M="M"
        "TM"        :@{->}_{p} "M" 
  }
\]
Here $\fake$ is the {\bf fake tangent bundle} or {\bf internal space}---a bundle over spacetime $M$ which is isomorphic to the tangent bundle $TM$, but also equipped with a fixed metric $\eta$.   The name coframe field comes from the case where $TM$ is trivializable, and $e\maps TM \to \fake = M\times \R^{3,1}$ is a choice of trivialization.  In this case $e$ restricts to a coframe $e_x\maps T_x M \to \R^{3,1}$ on each tangent space.  In any case, since $\fake$ is {\em locally} trivializable, we can treat $e$ locally as an $\R^{3,1}$-valued 1-form.

The tangent bundle acquires a metric by pulling back the metric on $\fake$:
\[
          g(v,w) := \eta(ev,ew)
\]
for any two vectors in the same tangent space $T_x M$.  In index notation, this becomes  $g_{\xa\xb}  = e^a_\xa  e^b_\xb \eta_{ab}$. In the case where the metric $g$ corresponds to a classical solution of general relativity, $e\maps TM\to \fake$ is an {\em isomorphism}, so that $g$ is nondegenerate.  However, the formalism makes sense when $e$ is any bundle morphism, and it is arguable whether one should allow degenerate coframe fields when attempting path-integral quantization.

When $e$ is an isomorphism, we can also pull a connection $\om$ on the vector bundle $\fake$ back to a connection on $TM$ as follows.  Working in coordinates, the covariant derivative of a local section $s$ of $\fake$ is 
\[
       (D_\mu s)^a = \partial_\mu s^a + \om^{a}_{\mu b} s^b
\]
When $e$ is an isomorphism, we can use $D$ to differentiate a section $w$ of $TM$ in the obvious way: use $e$ to turn $w$ into a section of $\fake$, differentiate this section, and use $e^{-1}$ to turn the result back into a section of $TM$.  This defines a connection on $TM$ by:
\[
     \del_v w = e^{-1} D_v  ew
\]
for any vector field $v$.  In particular, if $v=\partial_\mu$, $\del_\mu := \del_{\partial_\mu}$, we get, in index notation:
\begin{align*} 
   (\del_\mu w)^\xa &= 
             \partial_\mu  w^\xa
               + \Gamma^\xa_{\mu\xb} w^\xb
\end{align*}
where 
\begin{align*}
   \Gamma^\xa_{\mu\xb} :%&= e^\xa_a\partial_\mu e^a_\xb  + e^\xa_aA^{a}_{\mu b}  e^b_\xb \\
                                             &= e^\xa_a(\delta^a_b \partial_\mu   + \om^{a}_{\mu b})  e^b_\xb
\end{align*}

 The Palatini action is
\beq
\label{palatini}
S_{\rm Pal}(\om,e) =   \frac{1}{2G}  \int_M \star \left(e\wedge e\wedge R - \frac{\Lambda}{6}\, e\wedge e\wedge e\wedge e \right).
\eeq
where $R$ is the curvature of $\om$ and the wedge product $\wedge$ acts on both spacetime indices and internal Lorentz indices.  Compatibility with the metric $\eta$ forces the curvature $R$ to take values in $\Lambda^2\fake$.  Hence, the expression in parentheses is a $\Lambda^4\fake$-valued 4-form on $M$.  The $\star$ is an internal Hodge star operator, which turns such a form into an ordinary real-valued 4-form using the volume form and orientation on the internal space $\fake$:
\[
     \star\maps \Omega(M,\Lambda^4\fake) \to \Omega(M,\R)
\]
With internal indices written explicitly, this action is:
\[
S_{\rm Pal}(\om,e) =   \frac{1}{2G}  \int_M \left(e^a \wedge e^b \wedge R^{cd} 
        - \frac{\Lambda}{6} e^a \wedge e^b\wedge e^c \wedge e^d \right)\varepsilon_{abcd}.
\]
The variations of $\om$ and $e$ give us the respective equations of motion 
\begin{gather}
        d_\om(e\wedge e) = 0\label{pal-tor}\\
        e \wedge R - \frac{\Lambda}{3}\,  e\wedge e\wedge e = 0.  \label{pal-einstein}  
\end{gather}
In the classical case where $e$ is an isomorphism, the first of these equations is equivalent to 
\[
       d_\om e= 0
\]
which says precisely that the induced connection on $TM$ is torsion free, hence that $\Gamma^\xa_{\mu\xb}$ is the Christoffel symbol for the Levi-Civita connection. The other equation of motion, rewritten in terms of the metric and Levi-Civita connection,  is Einstein's equation.

\subsection{The coframe field}
\label{sec:coframe}

In describing Cartan geometry, what I called the `coframe field' was a nondegenerate, $H$-equivariant 1-form on the total space of the bundle $P$, with values in $\g/\h$:
\[
    e\maps TP \to \g/\h
\]
This is superficially quite different from the coframe field  
\[
     e\maps TM \to \fake
\]
used in the Palatini formulation of \GR.  The latter is a $\fake$-valued 1-form on spacetime; the former is a 1-form not on spacetime $M$, but on some principal bundle $P$ over $M$, with values not in a vector bundle, but in a mere vector space $\g/\h$.  To understand Palatini gravity in terms of Cartan geometry, we must see how these are really two descriptions of the same field. 

We first note that from the Cartan perspective, there is a natural choice of fake tangent bundle $\fake$.  Namely, for a Cartan geometry $P\to M$ modeled on $G/H$, the tangent bundle is isomorphic to the bundle associated to $P$ via the quotient representation of $\Ad(H)$ on $\g/\h$, so we take this as the {\bf fake tangent bundle}:
\[
        \fake := FM \times_H \g/\h.
\]
This is isomorphic, as a vector bundle, to the tangent bundle $TM$, but is equipped with a metric induced by the metric on $\g/\h$, provided the model geometry is a {\em metric} Klein geometry.  As explained below, with this choice of $\fake$, the two versions of the `coframe field' 
are in fact equivalent ways of describing the same field, given an Ehresmann connection on $P$.  In the reductive case---including the six model spacetimes we have considered---we get such an Ehresmann connection by projecting into the subalgebra $\h$, so in this case there is a canonical correspondence between the two descriptions of the coframe field.

To prove this correspondence, first suppose we have an Ehresmann connection $\omega$ on a principal $H$-bundle $p\maps P\to M$, and a Lie algebra $\g\supset \h$.
Given a 1-form $e\maps TM \to P\times_H \g/\h$ valued in the associated bundle, we wish to construct an $H$-equivariant 1-form $\tilde e\maps TP \to \g/\h$.   For any $v\in T_yP$, taking $e(d\pi(v))$ gives an element $[y',X]\in P\times_H \g/\h$.  This element is by definition an equivalence class such that $[y',X]=[y'h,\Ad(h^{-1})X]$ for all $h\in H$.   We thus define $\tilde e(v)$ for $v\in T_yP$ to be the unique element of $\g/\h$ such that $e(d\pi(v)) = [y,\tilde e(v)]$. 
This construction makes $\tilde e$ equivariant with respect to the actions of $H$, since on one hand 
\[
   e(d\pi(v)) = [y, \tilde e(v)] = [yh, \Ad(h^{-1})\tilde e (v)],
\]
while on the other
\[
   e(d\pi(v)) = e(d\pi({R_h}_\ast v)) = [yh, \tilde e({R_h}_\ast v)] = [yh, R_h^\ast \tilde e( v)],
\] 
so that
\[
R_h^\ast \tilde e( v) = \Ad(h^{-1})\tilde e (v).
\]

Conversely, given the equivariant 1-form $\tilde e\maps TP \to \g/\h$, define $e\maps TM \to P\times_H \g/\h$ as follows.  If $v\in T_xM$, pick any $y\in p^{-1}(x)$ and let $\tilde v_y \in T_yP$ be the unique horizontal lift of $v$ relative to the connection $\omega$.  Then let $e(v)=[y,\tilde e(\tilde v_y)]\in FM\times_H \g/\h$.  This is well-defined, since for any other $y' \in p^{-1}(x)$, we have $y'=yh$ for some $h\in H$, and hence
\[
     [y',\tilde e(\tilde v_{y'})]= [yh,R_h^\ast \tilde e(v_y)] = [yh,\Ad(h^{-1})\tilde e(v_y)] = [y,\tilde e(v_y)]
\]
where the second equality is equivariance and the third follows from the definition of the associated bundle $FM\times_H \g/\h$.  It is straightforward to show that the construction of $\tilde e$ from $e$ and vice-versa are inverse processes, so we are free to regard the coframe field $e$ in either of these two ways.
 
As mentioned in the previous section, for applications to quantum gravity it may be desirable to allow degenerate coframe fields, which don't correspond to classical solutions of general relativity.  The remarks of this section still hold for possibly degenerate coframe fields, provided we replace the Cartan connection with a generalized Cartan connection, as defined in Section~\ref{sec:exact}.

\subsection{\MM\ gravity}
\label{sec:MM}

Using results of the previous section, the Palatini action for general relativity can be viewed in terms of Cartan geometry, simply by thinking of the coframe field and connection as parts of a Cartan connection $A=\om+e$.  However, in its usual form:
\[
     S_{\rm Pal} = \frac{1}{2G}  \int_M \left(e^a \wedge e^b \wedge R^{cd} 
        - \frac{\Lambda}{6} e^a \wedge e^b\wedge e^c \wedge e^d \right)\varepsilon_{abcd}
\]
the action is not written directly in terms of the Cartan connection.  The \MM\ action can be seen as a rewriting of the Palatini action that makes the underlying Cartan-geometric structure more apparent.  

To obtain the \MM\ action, our first step will be to rewrite the Palatini action (\ref{palatini}) to look more like a gauge theory for the Lorentz group.  We can use the isomorphism $\Lambda^2\R^{3,1} \iso \so(3,1)$ to think of both $R$ and $e\wedge e$ as $\so(3,1)$-valued 2-forms.  We can also think of the $\varepsilon_{abcd}$ in the action as $-2!{\star}$, where $\star$ now denotes the Hodge star operator $\so(3,1)$ inherits from $\Lambda^2\R^{3,1}$.  We can then write the Palatini action as
\beq
\label{palstar}
     S_{\rm Pal}= \frac{-1}{G}  \int \tr \left((e\wedge e \wedge \star R 
        - \frac{\Lambda}{6} e \wedge e \wedge {\star(e \wedge e)}\right).
\eeq
where the trace is simply the matrix trace in the fundamental representation of $\so(3,1)$, as used in the metric (\ref{killing}) on the Lie algebra.  The action now resembles the Yang--Mills inner product of fields, $\int F\wedge {\ast F}$, except that the spacetime Hodge star $\ast$ has been replaced with the internal star. 
 
In each of the models discussed in Section~\ref{sec:modelspacetimes}, the $\so(3,1)$ or $\so(4)$ part of the curvature of the reductive Cartan connection, with appropriate internal cosmological constant, is given by
\[
       \widehat F = R - \frac{\xe}{\ell^2} e \wedge e,
\]
When $\Lambda\neq 0$, this gives us an expression for $e\wedge e$ which can be substituted into the Palatini action (\ref{palstar}) to obtain
\begin{align*}
       S &= \frac{-1}{G}  \int \tr \left( \frac{3}{\Lambda}(R-\widehat F) \wedge {\star R} - \frac{3}{2\Lambda}(R-\widehat F)\wedge {\star (R- \widehat F)} \right) \\
         &= \frac{-3}{2G\Lambda} \int \tr \left(\widehat F \wedge{\star \widehat F} + R\wedge {\star R}\right). 
\end{align*}
The $R\wedge {\star R}$ term here is a topological invariant, having vanishing variation due to the Bianchi identity.  This topological term is relevant for quantization, but for classical purposes, we may discard it,   
obtaining the {\bf MacDowell--Mansouri action} (\ref{MM-action}):
\begin{equation*}
  S_{\rm \scriptscriptstyle MM} = \frac{-3}{2G\Lambda} \int \tr(\widehat F \wedge \star \widehat F) 
\end{equation*}

The $BF$ reformulation of \MM\ gravity introduced by Freidel and Starodubtsev is given by the action
\[
       S =  \int \tr \left( 
              B\wedge F - \frac{\alpha}{2}\widehat B \wedge  {\star \widehat B}
              \right).
\]
where
\[
       \alpha = \frac{G\Lambda}{3}
\]
In deriving the classical field equations from this action, it is helpful to note that $\widehat B \wedge {\star \widehat B} = B \wedge {\star \widehat B}$.  Calculating the variation, we get:
\begin{align*}
  \xd S &=  \int \tr (\xd B \wedge (F - {\xa} {\star \widehat B}) + B\wedge \xd F ) \\
            &=  \int \tr (\xd B \wedge (F - {\xa} {\star \widehat B}) + d_A B\wedge \xd A )
\end{align*}
where in the second step we use the identity $\xd F = d_A \xd A$ and integration by parts.  The equations of motion resulting from the variations of $B$ and $A$ are thus, respectively,
\begin{align}
\label{FSmotion1}
F &= {\xa} {\star \widehat B} \\
\label{FSmotion2}
 d_A B &=0
\end{align}
Equivalently, we can decompose the $F$ and $B$ fields into reductive components, and rewrite these equations of motion as:
\begin{align}
\label{bfmm}
\begin{array}{rl}
\displaystyle
   R - \frac{\xe}{\ell^2} e \wedge e &= G\Lambda \star\!  \bs   \\
\displaystyle
   d_\om e &= 0 \\
\displaystyle
    d_\om  \bs + \frac{\xe}{\ell^2}(\bt\wedge e - e \wedge \bt) &= 0 \\
\displaystyle   
   d_\om \bt - \frac{1}{\ell} \bs\wedge e &= 0 
\end{array}
\end{align}
If we set $G=0$, these are identical to the equations for Cartan-type $BF$ theory obtained in Section~\ref{sec:BF}.  This means turning off Newton's gravitational constant turns 4d gravity into 4d Cartan-type $BF$ theory. 

Why are these the equations of general relativity?  Freidel and Starodubtsev approach this question indirectly, by solving (\ref{FSmotion1}) for $B$ and substituting back into the Lagrangian.  Doing this, and noting that $\star^2 = -1$, we obtain
\begin{align*}
     S &= \int \tr ( -\frac{1}{\xa} {\star F} \wedge \widehat F - \frac{1}{2\xa} {\star \widehat F} \wedge F ) \\
        &= \frac{-3}{2G\Lambda} \int \tr ( \widehat F \wedge {\star \widehat F})
\end{align*}
which is precisely the \MM\ action.

However, it is interesting to see Einstein's equations coming directly 
from the equations of motion (\ref{FSmotion1}) and (\ref{FSmotion2}).  For this, let us use the equivalent equations (\ref{bfmm}) in terms of constituent fields. Taking the covariant differential of the first equation shows, by the Bianchi identity $d_\om R=0$ and the second equation of motion---the vanishing of the torsion $d_\om e$---that
\[
       d_\om {\star \bs} = 0.
\]
But this covariant differential passes through the Hodge star operator, as shown in the Appendix, and hence 
\[
     d_\om \bs=0.
\]
This reduces the third equation of motion to
\[
   e^a \wedge \bt^b = e^b \wedge \bt^a.
\]
The matrix part of the form $e\wedge \bt$ is thus a {\em symmetric} matrix which lives in $\Lambda^2\R^4$, hence is zero.  When the coframe field $e$ is invertible, we therefore get
\[
        \bt = 0
\] 
and hence by the fourth equation of motion,
\[
    \bs\wedge e = 0.
\]
This in turn implies $e\wedge {\star \bs} = 0$, so wedging the first equation of motion with $e$ gives:
\[
 e\wedge (R - \frac{\xe}{\ell^2} e\wedge e) = 0.
\]
Using the appropriate cosmological constant (\ref{internal}), this is just Einstein's equation (\ref{pal-einstein}).

\section{Conclusions and Outlook}

In this paper, I have made a case for Cartan geometry as a means to a deeper understanding of the geometry of general relativity, particularly in the \MM\ formulation.  Cartan geometry not only gives geometric meaning to the \MM\ formalism (and to related theories \cite{symmetric}), but also deepens the connection between 4d $BF$ theory and gravity.  There are many unanswered questions raised by this work, which I must leave to future research.  I list just a few issues.

First, there seem to be many interesting questions regarding the $\Lambda \to 0$ limit of \MM\ gravity.  On one hand, the $BF$ formulation \cite{FreidelStarodubtsev}:
\[
       S =  \int \tr \left(B\wedge F - \frac{G\Lambda}{6}\widehat B \wedge  {\star \widehat B}\right).
\]
appears to be a `perturbation' of gravity around a $BF$ theory with gauge group $\SO(4,1)$ (or $\SO(3,2)$ or $\SO(5)$).  From the perspective I have presented here, however, in taking the $\Lambda \to 0$ limit, it seems natural to simultaneously let the `internal' cosmological constant of the Kleinian model tend to zero.     In this limit, the de Sitter or anti de Sitter group undergoes a Wigner contraction to the Poincar\'e group, so from the Cartan-geometric perspective, the $\Lambda\to 0$ limit of \MM\ gravity should be \MM\ theory with gauge group $\ISO(3,1)$.  This is also a topological theory, but it is a different topological theory from $\SO(4,1)$ $BF$ theory.  In fact, the \MM\ action (\ref{MM-action}) for $G = \ISO(3,1)$ reduces to an $R\wedge R$ theory for the Lorentz group.   

Another issue is that, given the relationship between `doubly special relativity' and de Sitter spacetime \cite{KG}, it is interesting to wonder whether, from the Cartan perspective, deformed special relativity might show up naturally as a limit of \MM\ gravity.  In fact, some new work by Gibbons and Gielen uses the Cartan geometric approach I have presented here to generalize from deformed special relativity to `deformed general relativity' \cite{GibbonsGielen}. 

It is also interesting to consider describing matter in quantum gravity, by first studying matter in $BF$ theory and then using the $BF$ reformulation of \MM\ gravity.  Some work has already been done in this direction \cite{loopbraid, BaezPerez, FKGS}, but the switch to {\em Cartan}-type $BF$ theory may have important consequences for this effort, which should be investigated.  

Finally, study of gauge theory based on Cartan connections may shed some light on an important problem in quantum gravity.  Naively, at least, gravity should have a constraint that says the coframe field (or the metric) is nondegenerate.  But $\det e \neq 0$ is not an {\em equation}, so this constraint can not be imposed by standard field theory methods.  One can argue that degenerate coframe fields might be important in quantum gravity, but the bottom line is that we do not know how to impose this constraint, even if we should!  From the Cartan perspective, however, this nondegeneracy constraint is precisely the condition for a certain Ehresmann connection to be a Cartan connection.  In other words, it is the condition for a connection to describe `rolling without slipping'. Moreover, since we can write down Cartan-type gauge theories that are much simpler than \MM\ gravity---including {\em topological} Cartan gauge theories---this may give a nice way to study this problem on its own, without the additional complications intrinsic to gravity.  

\subsection*{Acknowledgments}
I thank John Baez, Aristide Baratin, Jim Dolan, Laurent Freidel, Bill Goldman, Jeff Morton, Artem Starodubtsev, Danny Stevenson, and Josh Willis for helpful discussions.
I also thank Garrett Lisi and Jim Stasheff for catching errors in the first arxiv version.  Revisions of this manuscript were supported in part by NSF grant DMS-0636297.

\subsection*{Appendix: Internal Hodge star}
%\addtocontents{toc}{
%\contentsline {section}{\numberline {}Appendix: Hodge duality in $\so(3,1)$}{\thepage}}

The Hodge duals in this paper use the following conventions, for the exterior algebra $\Lambda V$ of an $n$-dimensional vector space $V$ with inner product $\eta$ of signature
\[
(\;\underbrace{{-}\cdots {-}}_{s}\underbrace{{+}\cdots {+}}_{n-s}\;).
\]
Letting $\{\xi_i \; |\; i=1, \ldots, n\}$ be an ordered orthonormal basis for $V$, we normalize the Hodge star operator so that
\[
        \star(\xi_1 \wedge \cdots \wedge \xi_p) = \xi_{p+1} \wedge \cdots \wedge \xi_n. 
\]
for $p$ from $0$ to $n$. Writing an arbitrary element $\om\in \Lambda^p V$ as
\[
       \om = \frac{1}{p!} \om_{i_1\cdots i_p} \xi^{i_1} \wedge \cdots \wedge \xi^{i_p}
\]
with antisymmetric components $\om_{i_1\cdots i_p} \xi^{i_1}$, this implies the Hodge dual $\star\om\in \Lambda^{n-p} V$:
\[
     \star\om = \frac{1}{(n-p)!} {\star\om}_{j_1 \cdots j_{n-p}} \xi^{j_1} \wedge \cdots \wedge \xi^{j_{n-p}}
\]
has components given by
\[
     \star\om_{j_1 \cdots j_{n-p}} =  \frac{1}{p!}  {\varepsilon^{i_1\cdots i_p}}_{j_1 \cdots j_{n-p}}   \om_{i_1\cdots i_p}
\]
The Hodge star acting on $p$-forms satisfies
\[
        \star^2 = (-1)^{p(n-p)+s}.
\]

The 6-dimensional Lie algebra $\so(3,1)$ inherits a notion of Hodge duality by the fact that it is isomorphic as a vector space to $\Lambda^2 \R^4$:
\[
\xygraph{
 []!{0;<3cm,0cm>:<-1.5cm,.5cm>::}
[] {\so(3,1)}="A" :@//^{} [r] {\Lambda^2 \R^4} 
                                       :@//^{\star} [r] {\Lambda^2 \R^4}
                                       :@//^{} [r] {\so(3,1)} 
"A" [d] {\text{\footnotesize lower an index}}  
 [r] {\text{\footnotesize Hodge duality}}
 [r] {\text{\footnotesize raise an index}}
}
\]
Using the above conventions on Hodge duals, one can check that the Hodge star permutes $\so(3,1)$ matrix entries, in the fundamental representation, as follows: 
\[
\star\left[
   \begin{array}{rrrr}
      0& a& b& c\\
      a& 0& d& e\\
      b& -d& 0& f\\
      c& -e& -f& 0\\
   \end{array}
\right]
=
\left[
   \begin{array}{rrrr}
      0& -f& e& -d\\
      -f& 0& c& -b\\
      e& -c& 0& a\\
     -d& b& -a& 0\\
   \end{array}
\right]
\]
It is then straightforward to verify the following properties of $\star$ on $\so(3,1)$:
\begin{itemize}
\item $\star\star X = -X$
\item $\star [X,X']=[X,\star X']$
\end{itemize}
 For \MM\ gravity and its $BF$ reformulation, the essential application of the second property is that if $\om$ is an $\SO(3,1)$ connection, the covariant differential $d_\om$ commutes with the internal Hodge star operator:
\[
   d_\om (\star X) = d (\star X) + [\om,\star X]  = \star(dX + [\om,X]) = \star d_\om X.
\]

The $\so(4)$ case, relevant for \MM\ gravity in Riemannian signature, is similar, the main difference being that $\star^2= +1$.

\end{document}